\documentclass[11pt]{article}
\usepackage{amssymb}
\usepackage{amsmath}
\usepackage{amstext}
\usepackage{graphicx,epsfig}
\usepackage{epsfig}
\usepackage{verbatim} 
\usepackage{empheq,fancybox}
\usepackage{color}
\usepackage{ulem}
\usepackage{enumitem}
\usepackage{subfigure}
\usepackage{bbm}
\usepackage{parskip}
\usepackage{dsfont}
\usepackage[numbers,sort&compress]{natbib}
\usepackage{bm}
\usepackage[dvipsnames]{xcolor}
\usepackage{multirow}
\usepackage{cancel}
\usepackage{tensor}

\usepackage{hyperref} %Automatically links \label and \ref commands; Always load last
\usepackage[all]{hypcap} %Link navagates to top of figure instead of caption (below fig)
\hypersetup{
    colorlinks=true,       % false: boxed links; true: colored links
    linkcolor=red,          % color of internal links
    citecolor=blue,        % color of links to bibliography
    filecolor=magenta,      % color of file links
    urlcolor=blue           % color of external links
}

\usepackage{myterms}

\newcommand{\CME}{\text{\sc \fontfamily{ptm}\selectfont cme}}

\newcommand{\Ee}{\text{\sc e}{\rm e}}

\newcommand{\Nhe}{\nu{\rm he}}
\newcommand{\Ehe}{\text{\sc e}{\rm he}}

\newcommand{\CS}{\text{\sc cs}}
\newcommand{\EW}{\text{\sc ew}}
\newcommand{\TS}{\text{\sc ts}}
\renewcommand{\Gauss}{\ \mathrm{G}}

\newcommand{\tW}{\theta_{\text{\sc \fontfamily{ptm}\selectfont w}}}
\newcommand{\tWz}{\theta_{\text{\sc \fontfamily{ptm}\selectfont w}0}}
\newcommand{\dtW}{\frac{d\tW}{dt}}

\newcommand{\MHD}{\text{\sc mhd}}

\numberwithin{equation}{section}

\begin{document}
%
%\maketitle
~
\vspace{1truecm}
\renewcommand{\thefootnote}{\fnsymbol{footnote}}
\begin{center}
{\huge \bf{
Evolution of the Baryon Asymmetry 
\\ \vspace{0.0cm} 
through the Electroweak Crossover
\\ \vspace{0.3cm} 
in the Presence of a Helical Magnetic Field 
}}
\end{center} 

\vspace{1truecm}
\thispagestyle{empty}
\centerline{\Large Kohei Kamada${}^{\rm a}$\footnote{\tt kohei.kamada@asu.edu} and Andrew J. Long${}^{\rm b}$\footnote{\tt andrewjlong@kicp.uchicago.edu}}
\vspace{.7cm}

\centerline{\it ${}^{\rm a}$School of Earth and Space Exploration, Arizona State University,
Tempe, AZ 85287, USA.}

\vspace{.2cm}
\centerline{\it$^{\rm b}$Kavli Institute for Cosmological Physics, University of Chicago, Chicago, Illinois 60637, USA}

\vspace{.5cm}
\begin{abstract}
\vspace{.03cm}
\noindent

%================
We elaborate upon the model of baryogenesis from decaying magnetic helicity by focusing on the evolution of the baryon number and magnetic field through the Standard Model electroweak crossover.  
The baryon asymmetry is determined by a competition between the helical hypermagnetic field, which sources baryon number, and the electroweak sphaleron, which tends to wash out baryon number.  
At the electroweak crossover both of these processes become inactive;  the hypermagnetic field is converted into an electromagnetic field, which does not source baryon number, and the weak gauge boson masses grow, suppressing the electroweak sphaleron reaction.  
An accurate prediction of the relic baryon asymmetry requires a careful treatment of the crossover.  
We extend our previous study [K.~Kamada and A.~J.~Long, Phys.~Rev.~D~{\bf 94},~065301~(2016)], taking into account the gradual conversion of the hypermagnetic into the electromagnetic field.  If the conversion is not completed by the time of sphaleron freeze-out, as both analytic and numerical studies suggest, the relic baryon asymmetry is enhanced compared to previous calculations.  
The observed baryon asymmetry of the Universe can be obtained for a primordial magnetic field that has a present-day field strength and coherence length of $B_0 \sim 10^{-17} \, {\rm G}$ and $\lambda_0 \sim 10^{-3} \, {\rm pc}$ and a positive helicity.  
For larger $B_0$ the baryon asymmetry is overproduced, which may be in conflict with blazar observations that provide evidence for an intergalactic magnetic field of strength $B_0 \gtrsim 10^{-14\sim16} \, {\rm G}$.  

\end{abstract}

\newpage

\begingroup
\hypersetup{linkcolor=black}
\tableofcontents
\endgroup

\renewcommand*{\thefootnote}{\arabic{footnote}}
\setcounter{footnote}{0}

%==================================
% INTRODUCTION
%==================================
\section{Introduction}\label{sec:Introduction}

%================
The origin of the matter/antimatter asymmetry of the Universe [or baryon asymmetry of the Universe (BAU)] remains a long-standing problem at the interface of cosmology and high energy physics.  
In order to generate a baryon asymmetry from an initially matter/antimatter symmetric universe, it is necessary for the system to contain processes that violate baryon number \cite{Sakharov:1967dj}. 
Such processes are already present in the Standard Model (SM) due to field theoretic quantum anomalies \cite{Adler:1969gk,Bell:1969ts,Hooft:1976up}. 
These anomalous processes involve either the $\SU{2}_L$ weak isospin gauge fields or the $\U{1}_Y$ hypercharge gauge field.  
Whereas SM baryon-number violation via the $\SU{2}_L$ gauge field features prominently in many models of baryogenesis, such as electroweak baryogenesis and leptogenesis, we are interested in SM baryon-number violation via the $\U{1}_Y$ gauge field.  

%================
In the symmetric phase of the electroweak (EW) plasma ($T \gtrsim 160 \GeV$ in the SM \cite{Kajantie:1996mn}), the anomaly expresses the fact that changes in baryon number ($Q_B$) and lepton number ($Q_L$) can be induced by changes in $\SU{2}_L$ Chern-Simons number ($N_{\CS}$) or $\U{1}_{Y}$ hypermagnetic helicity ($\Hcal_{Y}$) as 
\begin{align}\label{eq:Bdot}
	\Delta Q_{B} =\Delta Q_{L} = N_{\rm g} \Delta N_{\CS} - N_{\rm g} \frac{g^{\prime 2}}{16\pi^2} \Delta \Hcal_{Y}
	\per
\end{align}
The factor of $N_{\rm g} = 3$ is the number of fermion generations and $g^{\prime}$ is the $\U{1}_Y$ gauge coupling.  
Thermal fluctuations of the $\SU{2}_L$ gauge fields (EW sphalerons \cite{Rubakov:1996vz}) allow $N_{\CS}$ to diffuse, which pushes $Q_B$ and $Q_L$ to zero (assuming a vanishing $B-L$ asymmetry).
The system may also contain a helical hypermagnetic field, {\it i.e.} a primordial magnetic field~(PMF) in the symmetric phase of the EW plasma associated with $\U{1}_Y$ hypercharge that has  excess power in either the left- or right-circular polarization mode.  
A helical PMF can arise, for example, from axion dynamics during inflation \cite{Turner:1987bw, Garretson:1992vt, Anber:2006xt, Durrer:2010mq, Jain:2012jy, Adshead:2015pva,Fujita:2015iga, Adshead:2016iae} (see also Refs.~\cite{DiazGil:2007dy,DiazGil:2008tf}).
Due to interactions of the hypermagnetic field with the charged plasma, the hypermagnetic helicity slowly decays.  
If $\Hcal_{Y} > 0$ initially, then $\Delta \Hcal_{Y} < 0$ implies the generation of a baryon asymmetry $\Delta Q_B > 0$.  
In this way, the BAU may have arisen from a helical hypermagnetic field in the early Universe.  

%================
Various studies have explored the relationship between baryon-number violation and magnetic fields in the early Universe.  
Among the earliest works, Joyce and Shaposhnikov~\cite{Joyce:1997uy} showed that a helical hypermagnetic field can arise in the symmetric phase of the EW plasma from a preexisting lepton asymmetry carried by the right-chiral electron \cite{Campbell:1992jd} (see also Refs.~\cite{Long:2013tha,Long:2016uez}).
This work was soon extended by Giovannini and Shaposhnikov \cite{Giovannini:1997eg,Giovannini:1997gp,Giovannini:1999by,Giovannini:1999wv} to consider the generation of baryon-number isocurvature fluctuations from a preexisting stochastic hypermagnetic field.  
These ideas were formulated into a model of baryogenesis by Bamba~\cite{Bamba:2006km} where the dynamics of an axion field during inflation leads to the growth of a helical hypermagnetic field with a large correlation length, which is partially converted into baryon number by the SM anomalies at the electroweak phase transition (see also Refs.~\cite{Bamba:2007hf,Anber:2015yca}).  
Other related work has explored the connection between helical magnetic fields in the early Universe and the anomalous violation of chiral charge \cite{Boyarsky:2011uy,Boyarsky:2012ex,Zadeh:2015oqf,Boyarsky:2015faa,Gorbar:2016qfh} (see also Refs.~\cite{Giovannini:2013oga,Giovannini:2015aea,Giovannini:2016whv}) and lepton number \cite{Semikoz:2003qt,Semikoz:2004rr,Semikoz:2005ks,Semikoz:2009ye,Akhmet'ev:2010ba,Dvornikov:2011ey,Semikoz:2012ka,Dvornikov:2012rk,Semikoz:2013xkc,Semikoz:2015wsa,Semikoz:2016lqv}.  

%================
Models of baryogenesis from magnetogenesis are interesting in part because the primordial magnetic field is expected to persist today as an intergalactic magnetic field (IGMF).  
Although the existence and origin of the IGMF have not yet been established, the body of evidence is growing.  
(See Refs.~\cite{Durrer:2013pga,Subramanian:2015lua} for recent reviews on cosmological magnetic fields.)
Recent measurements of TeV blazar spectra have identified a deficit of secondary cascade photons.  
These observations can be explained to result from a magnetic broadening of the cascade by the IGMF \cite{Neronov:1900zz,Tavecchio:2010mk,2011ApJ...727L...4D,Essey:2010nd,Taylor:2011bn,Takahashi:2013lba,Finke:2015ona}, which thereby provides indirect evidence for the existence of a PMF with a field strength and coherence length today of $B_0 \gtrsim 10^{-14\sim16} \Gauss$ and $\lambda_0 \gtrsim 10^{-2}\sim 1 \pc$.  
Similarly, searches for GeV pair halos around TeV blazars have also reported evidence for an IGMF \cite{Ando:2010rb,Chen:2014rsa} (see also Refs.~\cite{AlvesBatista:2016urk,Broderick:2016akd}).  
Additionally, measurements of the diffuse gamma ray flux at Earth suggest a parity-violating character in gamma ray arrival directions, which can be interpreted as evidence for the presence of a helical IGMF \cite{Tashiro:2013bxa,Tashiro:2013ita,Tashiro:2014gfa,Chen:2014qva}.  

%================
Motivated in part by these new probes of the IGMF, Fujita and Kamada~\cite{Fujita:2016igl} recently revisited baryogenesis from hypermagnetic helicity.  
By drawing on the results of recent magnetohydrodynamic simulations, they used an improved model for the evolution of the magnetic field (inverse cascade behavior) to calculate the slowly decaying magnetic helicity and corresponding production of baryon number.  
Their calculation indicates that a maximally helical magnetic field stronger than $B_0 \sim 10^{-12} \Gauss$ today would generate a much larger baryon number than what is observed.  
Since this baryogenesis is an inevitable consequence of SM physics once the helical hypermagnetic field is provided, there is a mild conflict between the observed BAU and blazar observations, which favor $B_0 \gtrsim 10^{-14\sim16} \Gauss$.  

%================
However, none of these studies directly addresses the conversion of the hypermagnetic field into an electromagnetic field at the EW crossover and the corresponding effect on the relic baryon asymmetry.  
Since the electromagnetic field has vectorlike interactions, it does not contribute to the baryon-number anomaly.  
Therefore, if the conversion completes before the EW sphalerons freeze-out, the sphalerons threaten to erase the baryon asymmetry.  
In the early works of Giovannini and Shaposhnikov and Bamba, {\it et al.}~\cite{Bamba:2006km,Bamba:2007hf,Giovannini:1997eg,Giovannini:1997gp,Giovannini:1999by,Giovannini:1999wv} it was argued that the EW phase transition must be first order so that the EW sphaleron process is out of equilibrium in the broken phase and washout of baryon number is avoided.  
The assumption is implicit in later work \cite{Anber:2015yca,Fujita:2016igl} where baryon-number violation due to both the EW sphaleron and the hypermagnetic field are assumed to shut off simultaneously at EW temperatures.  

%================
Kamada and Long~\cite{Kamada:2016eeb} recently demonstrated that a complete washout of baryon number is avoided even if there is no $B-L$ asymmetry and the EW phase transition is a continuous crossover as we expect in the SM.
Although the EW sphaleron remains in thermal equilibrium until $T \simeq 130 \GeV$ \cite{DOnofrio:2014kta} after the hypermagnetic field has been converted to an electromagnetic field, and therefore no longer sources baryon number, washout is avoided because the EM field sources chirality and inhibits the communication of baryon-number violation from the left-chiral to right-chiral fermions.  
To model the conversion of the hypermagnetic field into electromagnetic field at the EW crossover, \rref{Kamada:2016eeb} assumed that the transformation occurs abruptly at a fiducial temperature of $T = 160 \GeV$ where the Higgs condensate first starts to deviate from zero (see also \rref{Pavlovic:2016mxq}).
As discussed in \rref{Kamada:2016eeb}, this is a conservative approach;  since the electromagnetic field does not violate baryon number, this approximation can underestimate the relic baryon asymmetry if the conversion of the hypermagnetic field into the electromagnetic field is gradual.  

%================
In this work, we develop a more sophisticated treatment for the conversion of hypermagnetic field into electromagnetic field at the EW crossover.  
By drawing on analytic and lattice results we see that the hypermagnetic field is not fully converted into an electromagnetic field even at temperatures as low as $T = 140 \GeV$.  
Therefore, the source term from decaying magnetic helicity remains active, while the washout by EW sphalerons goes out of equilibrium.  
Consequently, we show that the relic baryon asymmetry can be greatly enhanced as compared to \rref{Kamada:2016eeb}.  
It is possible to generate the observed BAU from a maximally helical magnetic field that was generated prior to the EW crossover and has a strength and coherence length today of about $B_0 \sim 10^{-16\sim17} \Gauss$ and $\lambda_0 \sim 10^{-2\sim3} \pc$.  
If the magnetic field strength is larger, such as $B_0 \gtrsim 10^{-14\sim16} \Gauss$ suggested by blazar observations, the relic baryon asymmetry is generally overproduced.  
This presents a new constraint for models of magnetogenesis that rely on inflation or cosmological phase transitions prior to the EW epoch.  

%================
The rest of the paper is organized as follows.  
In \sref{sec:Derivation}, we generalize the calculation of \rref{Kamada:2016eeb} to allow for a gradual conversion of the hypermagnetic field into an electromagnetic field at the EW crossover.  
In \sref{sec:Analytic}, we present an analytic solution of the kinetic equations, which gives the equilibrium baryon-number abundance.  
In \sref{sec:Results}, we solve the kinetic equations numerically and compare with our analytic formula.  
We show how the relic baryon asymmetry depends on the field strength and coherence length today.  
We see that baryon number is overproduced for relatively large magnetic field strength, $B_0 \gtrsim 10^{-16} \Gauss$. 
In \sref{sec:Avoid}, we discuss ways to avoid the baryon overproduction while also accommodating the IGMF interpretation of blazar observations.  
Finally we conclude in \sref{sec:Conclusion} and point to directions for future work.  

%==================================
% DERIVATION OF SOURCE TERMS
%==================================
\section{Derivation of source terms}\label{sec:Derivation}

%================
In this section, we generalize our previous calculation in order to model the gradual conversion of the hypermagnetic field into an electromagnetic field.  
For definitions and notation, the reader is referred to \rref{Kamada:2016eeb}.  

%==========
First, let us recall what is the quantity of interest.  
In the presence of a helical magnetic field, SM quantum anomalies lead to the appearance of source terms in the kinetic equations for the various SM particle asymmetries.  
These source terms appear in the kinetic equation for fermion species $f$ in the following way \cite{Kamada:2016eeb}: 
\begin{equation}\label{eq:detaf_dx}
	\frac{d\eta_f}{dx} = c_{1,f} \, \Scal_{\rm y}^{\rm bkg} + c_{2,f} \, \Scal_{\rm w}^{\rm bkg} + c_{3,f} \, \Scal_{\rm yw}^{\rm bkg} + \cdots
	\per
\end{equation}
Here, $\eta_f=n_f/s$ is the particle number asymmetry in species $f$ divided by the entropy density of the cosmological plasma.  
We use the dimensionless temporal coordinate $x \equiv T/H = M_0/T$ where $T$ is the temperature of the cosmological plasma and $H = T^2/M_0$ with $M_0 \simeq 7.1 \times 10^{17} \GeV$ is the Hubble parameter at temperatures where the entire SM particle content is relativistic.
The coefficients of the source terms $c_{i,f}$ depend on the quantum numbers of $f$; see \rref{Kamada:2016eeb}.  
The dots ($\cdots)$ represent other interactions in which a fermion of species $f$ participates.
These include Yukawa interactions, EW and strong sphalerons, and weak interactions. 
The source terms $\Scal$ take the form (see Eq.~(2.44) of \rref{Kamada:2016eeb})
\begin{subequations}\label{eq:sources_1}
\begin{align}
	\Scal_{\rm w}^{\rm bkg} & = \frac{1}{2} \Bigl( \frac{1}{sT} \frac{1}{16\pi^2} \Bigr) \ g^{2} \ \overline{ \langle W_{\mu \nu}^{a} \rangle \langle \widetilde{W}^{a \mu \nu} \rangle } \\
	\Scal_{\rm y}^{\rm bkg} & = \Bigl( \frac{1}{sT} \frac{1}{16\pi^2} \Bigr) \ g^{\prime 2} \ \overline{ \langle Y_{\mu \nu} \rangle \langle \widetilde{Y}^{\mu \nu} \rangle } \\
	\Scal_{\rm yw}^{\rm bkg} & = 2 \Bigl( \frac{1}{sT} \frac{1}{16\pi^2} \Bigr) \ g g^{\prime} \ \overline{ \langle Y_{\mu \nu} \rangle \langle \widetilde{W}^{3 \mu \nu} \rangle } 
	\per
\end{align}
\end{subequations}
where $Y_{\mu \nu}$ and $W_{\mu \nu}^{a}$ are the field strength tensor of $\U{1}_Y$ hypercharge and $\SU{2}_L$ isospin, respectively, and $g^{\prime}$ and $g$ are their respective coupling parameters.  
The dual tensor is defined by $\tilde{F}^{\mu \nu} = \epsilon^{\mu \nu \rho \sigma} F_{\rho \sigma}/2$ with normalization $\epsilon^{0123}=1$. 
The angled brackets indicate thermal ensemble averaging, and the bar denotes volume averaging.  
In this section, we seek to evaluate these three sources.  

%==========
Now, let us recall how we modeled the gauge fields during the EW crossover in \rref{Kamada:2016eeb}.  
We assumed that the system passes abruptly from the symmetric phase to a broken phase as the temperature is lowered through $T_{\rm co} \simeq 162 \GeV$ in a similar way to \rref{Pavlovic:2016mxq}.  
This numerical value is taken from lattice studies of the EW crossover \cite{DOnofrio:2015mpa}.  
In the symmetric phase ($T > T_{\rm co}$), the non-Abelian $\SU{2}_L$ gauge field is screened due to its self-interactions  \cite{Gross:1980br}, and it is well known that the corresponding isomagnetic field vanishes (up to thermal fluctuations).  
Meanwhile, the $\U{1}_Y$ sector is assumed to carry a hypermagnetic field ${\bm B}_{Y}\xt$ generated by a magnetogenesis mechanism that occurred before the EW crossover.  
In the broken phase ($T < T_{\rm co}$), the Higgs condensate induces a mass for charged $W_{\mu}^{\pm}(x)$ and neutral $Z_{\mu}(x)$ gauge fields.  
We argued that the massive fields decay quickly, leaving only the massless electromagnetic field $A_{\mu}(x)$.  
We defined the electromagnetic field through the standard electroweak rotation, $A_{\mu} = \sin \tWz \, W_{\mu}^{3} + \cos \tWz \, Y_{\mu}$, where the vacuum weak mixing angle $\tWz$ is expressed in terms of the $\U{1}_Y$ and $\SU{2}_L$ and gauge couplings, $g^{\prime}$ and $g$, respectively, as $\tan \tWz = g^{\prime} / g$.  
This relation furnishes the matching condition ${\bm B}_{A}({\bm x},t_{\rm co}+\epsilon) = \cos \tWz \, {\bm B}_{Y}({\bm x},t_{\rm co}-\epsilon)$, which we used to relate the electromagnetic field just after the crossover $t_{\rm co} \equiv t(T=162 \GeV)$ to the hypermagnetic field just before the crossover.  

%==========
The approach described above is not correct in the following sense.  
During the EW crossover, the gauge fields acquire mass from both the Higgs condensate {\it and thermal effects in the plasma}.  
If the thermal effects could be neglected, then we would have four massless fields in the symmetric phase where the Higgs condensate is zero, and we would have one massless field in the broken phase.  
If we define the weak mixing angle as the parameter of the $\SO{2}$ matrix that diagonalizes the quadratic gauge field terms in the Lagrangian, then this approximation corresponds to an abrupt change from $\tan \tW = 0$ in the symmetric phase to $\tan \tW = g^{\prime} / g$ in the broken phase.  
However, this is not the case.\footnote{We are grateful to Mikhail Shaposhnikov for bringing this point to our attention.}  
As we have already mentioned above, the non-Abelian gauge fields $W_{\mu}^{a}(x)$ also acquire mass from their self-interactions in the plasma, which leads to the screening of isomagnetic fields.  
Consequently, the mixing angle $\tW(t)$ will change slowly with time while interpolating smoothly between its symmetric and broken phase limiting values.  
It continues to deviate appreciably from its zero-temperature value $\tan \tWz = g^{\prime} / g$ even at relatively low temperatures of $T = 140 \GeV$.  
This behavior is confirmed by analytic calculations \cite{Kajantie:1996qd} and recent numerical lattice simulations~\cite{DOnofrio:2015mpa}.  
We will study it quantitatively in \sref{sec:Results}.  

%==========
In light of the preceding discussion, we generalize our treatment of the gauge fields at the EW crossover as follows.  
At any time, the spectrum consists of three massive and one massless gauge field degrees of freedom.  
In general, the massless degree of freedom at time $t$ can be written as an $\SO{2}$ rotation of $W_{\mu}^{3}(x)$ and $Y_{\mu}(x)$ with parameter $\tW(t)$.  
In other words, $\tW(t)$ is defined as the rotation angle that projects the massless field degree of freedom onto the $\U{1}_Y$ field axis.  
As before, we assume that the massive fields are screened or decay away quickly compared to the time scale on which the baryon asymmetry evolves.\footnote{This assumption is confirmed with the following rough estimates.  Parametrically, the perturbative Z-boson decay width at temperature $T$ is given by $\Gamma_Z \sim (g^2 + g^{\prime 2})^{3/2} v(T)$ where $v(T)$ is the vacuum expectation value of the Higgs field at temperature $T$.  Comparing this decay rate with the Hubble expansion rate during the EW epoch, we have $\Gamma_Z/H \sim 10^{15} (T/100\GeV)^{-2} (v(T)/100 \GeV)$, which supports our assumption that the Z-field decays quickly.  We expect this general conclusion to be unchanged when thermal and nonperturbative effects are considered more carefully.  } 
Therefore, the field evolution can be modeled by the ansatz 
\begin{subequations}\label{eq:Ansatz}
\begin{align}
	\langle W_{\mu}^{1}(x) \rangle & = \langle W_{\mu}^{2}(x) \rangle = 0 \\
	\langle W_{\mu}^{3}(x) \rangle & = \sin \tW(t) \, \Acal_{\mu}(x) \\
	\langle Y_{\mu}(x) \rangle & = \cos \tW(t) \, \Acal_{\mu}(x) 
	\per
\end{align}
\end{subequations}
By requiring the three massive field degrees of freedom to vanish and their decay not to affect the evolution of the massless field degree of freedom, we have reduced the problem to a single degree of freedom as represented by the classical vector field $\Acal_{\mu}(x)$.  

\begin{figure}[t]
\hspace{0pt}
\vspace{-0in}
\begin{center}
\includegraphics[width=0.5\textwidth]{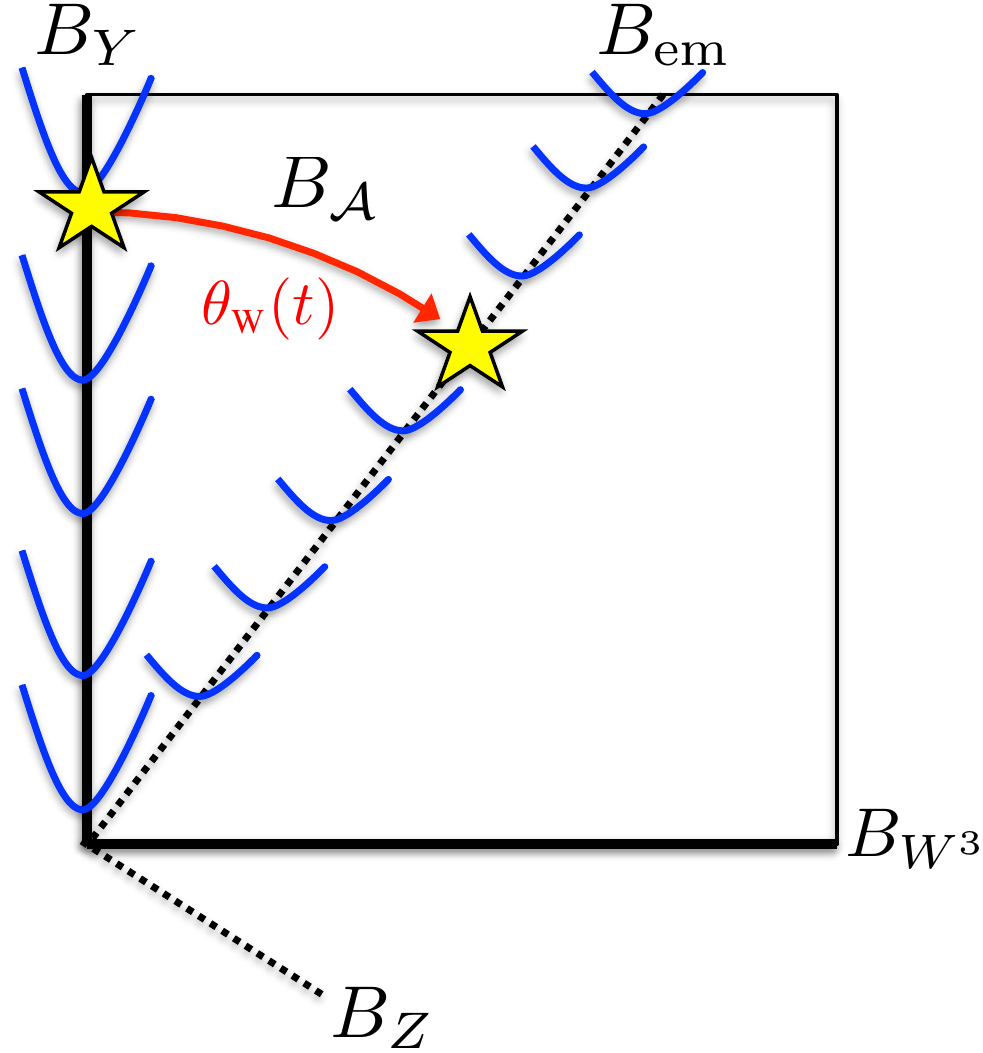}
\caption{
\label{fig:rotation_cartoon}
A graphical representation of the conversion from hypermagnetic field ${\bm B}_Y$ into electromagnetic field ${\bm B}_{\rm em}$ during the EW crossover.  The (blue) parabolas indicate the curvature of the thermal effective potential.  The weak mixing angle $\tW(t)$ measures the separation of the flat direction (massless field degree of freedom) and the $\U{1}_Y$ axis.
}
\end{center}
\end{figure}

%==========
The ansatz (\ref{eq:Ansatz}) is represented graphically in \fref{fig:rotation_cartoon}, which illustrates the conversion from hypermagnetic field to electromagnetic field.  
Here, we denote the magnetic field of a gauge field ${\cal Y}$ as ${\bm B}_{\cal Y} \equiv {\nabla} \times { {\cal Y}}$.  
We have drawn the figure so as to suggest that $|{\bm B}_{\Acal}|$ does not decrease appreciably during the EW crossover.  
As we will explain later, this is the case because $\Acal_\mu$ evolves slowly according to the cosmic expansion and the inverse cascade.  

%==================

%==========
Having generalized the gauge field ansatz from our earlier work, we are now prepared to revisit the calculation of source terms (\ref{eq:sources_1}).  
Using the ansatz in \eref{eq:Ansatz}, the source terms can be written as 
\begin{subequations}\label{eq:sources_2}
\begin{align}
	\Scal_{\rm w}^{\rm bkg} & = \frac{1}{2} \Bigl( \frac{1}{sT} \frac{1}{16\pi^2} \Bigr) \ g^{2} \Bigl( \sin^2 \tW(t) \overline{\Acal_{\mu \nu} \tilde{\Acal}^{\mu \nu}} + 2 \dtW \sin 2\tW(t) \tensor{\delta}{^{0}_{\! \mu}} \overline{ \Acal_{\nu} \tilde{\Acal}^{\mu \nu}} \Bigr) \\
	\Scal_{\rm y}^{\rm bkg} & = \Bigl( \frac{1}{sT} \frac{1}{16\pi^2} \Bigr) \ g^{\prime 2} \Bigl( \cos^2 \tW(t) \overline{\Acal_{\mu \nu} \tilde{\Acal}^{\mu \nu}} - 2 \dtW \sin 2\tW(t) \tensor{\delta}{^{0}_{\! \mu}} \overline{\Acal_{\nu} \tilde{\Acal}^{\mu \nu}} \Bigr) \\
	\Scal_{\rm yw}^{\rm bkg} & = 2 \Bigl( \frac{1}{sT} \frac{1}{16\pi^2} \Bigr) \ g g^{\prime} \Bigl( \sin \tW(t) \cos \tW(t) \overline{\Acal_{\mu \nu} \tilde{\Acal}^{\mu \nu}} + 2 \dtW \cos 2 \tW(t) \tensor{\delta}{^{0}_{\! \mu}} \overline{\Acal_{\nu} \tilde{\Acal}^{\mu \nu}} \Bigr) 
\end{align}
\end{subequations}
where $\Acal_{\mu \nu}$ is the field strength tensor associated with $\Acal_{\mu}(x)$ and ${\tilde \Acal}^{\mu \nu} = \epsilon^{\mu\nu\rho\sigma} \Acal_{\rho \sigma}/2$ is the dual tensor.  
In terms of the 3-vector notation, the two terms in parentheses are
\begin{align}
	\Acal_{\mu \nu} \tilde{\Acal}^{\mu \nu} = -4 {\bm E}_{\Acal} \cdot {\bm B}_{\Acal}
	\qquad \text{and} \qquad
	\tensor{\delta}{^{0}_{\! \mu}} \Acal_{\nu} \tilde{\Acal}^{\mu \nu} = {\bm A}_{\Acal} \cdot {\bm B}_{\Acal}
\end{align}
where ${\bm E}_{\Acal}\xt$ is the electric field with $\bigl( {\bm E}_{\Acal} \bigr)_i = \Acal_{0i}$, ${\bm B}_{\Acal}\xt$ is the magnetic field with $\bigl( {\bm B}_{\Acal} \bigr)_i = \tilde{\Acal}_{0i}$, and ${\bm A}_{\Acal}\xt$ is the vector potential with $\bigl( {\bm A}_{\Acal} \bigr)_{i} = \Acal^{i}$.  
With this replacement, the sources become
\begin{subequations}\label{eq:sources_3}
\begin{align}
	\Scal_{\rm w}^{\rm bkg} & = \frac{1}{2} \Bigl( \frac{1}{sT} \frac{1}{16\pi^2} \Bigr) \ g^{2} \Bigl( -4 \sin^2 \tW(t) \overline{ {\bm E}_{\Acal} \cdot {\bm B}_{\Acal}} + 2 \dtW \sin 2\tW(t) \overline{{\bm A}_{\Acal} \cdot {\bm B}_{\Acal}} \Bigr) \\
	\Scal_{\rm y}^{\rm bkg} & = \Bigl( \frac{1}{sT} \frac{1}{16\pi^2} \Bigr) \ g^{\prime 2} \Bigl( -4 \cos^2 \tW(t) \overline{{\bm E}_{\Acal} \cdot {\bm B}_{\Acal}} - 2 \dtW \sin 2\tW(t) \overline{{\bm A}_{\Acal} \cdot {\bm B}_{\Acal}} \Bigr) \\
	\Scal_{\rm yw}^{\rm bkg} & = 2 \Bigl( \frac{1}{sT} \frac{1}{16\pi^2} \Bigr) \ g g^{\prime} \Bigl( -4 \sin \tW(t) \cos \tW(t) \overline{{\bm E}_{\Acal} \cdot {\bm B}_{\Acal}} + 2 \dtW \cos 2 \tW(t) \overline{{\bm A}_{\Acal} \cdot {\bm B}_{\Acal}} \Bigr) 
	\per
\end{align}
\end{subequations}
The second term in parentheses is new, since we are now allowing $d\tW/dt \neq 0$.  
Recall that $\overline{{\bm A}_{\Acal} \cdot {\bm B}_{\Acal}}$ is the helicity of the gauge field $\Acal_{\mu}(x)$.  
Under a gauge transformation, we send ${\bm A}_{\Acal} \to {\bm A}_{\Acal} - {\bm \nabla} \chi$ ,and since ${\bm \nabla} \cdot {\bm B}_{\Acal} = 0$, the helicity density ${\bm A}_{\Acal} \cdot {\bm B}_{\Acal}$ transforms into itself up to a total 3-divergence.  
The volume averaged helicity $\overline{{\bm A}_{\Acal} \cdot {\bm B}_{\Acal}}$ is gauge invariant provided that the surface term vanishes; for example, see \rref{Brandenburg:2004jv}.  

%==========
To evaluate the electric field ${\bm E}_{\Acal}$, we recognize that the electric current ${\bm J}_{\Acal}$ is given by 
\begin{align}\label{eq:JA}
	{\bm J}_{\Acal} = \sigma_{\Acal} \bigl( {\bm E}_{A} + {\bm v} \times {\bm B}_{\Acal} \bigr) + {\bm J}_{\CME, \Acal}
	\per
\end{align}
The first term is simply Ohm's law with $\sigma_{\Acal}$ the conductivity.  
The second term is the chiral magnetic effect (CME) current, which we evaluate below.  
The current ${\bm J}_{\Acal}$ also appears in the equation of motion\footnote{
Here, we gloss over some subtleties related to gauge invariance.  In general, the transformation (\ref{eq:Ansatz}) should be generalized to include the orthogonal field direction $\Zcal_{\mu}(x)$.  Due to the time-dependent linear transformation, the field equations for $\Acal$ and $\Zcal$ acquire ``mass terms'' of the form $(d\tW/dt)^2 {\bm A}_{\Acal}$ and $(d\tW/dt)^2 {\bm A}_{\Zcal}$.  Nevertheless, one can verify explicitly that the field equations are gauge invariant.  This is because the field strength tensors are no longer invariant under the gauge transformation when $(d\tW/dt) \neq 0$.  Despite these subtleties, we have checked that the source terms appearing in \eref{eq:sources_3} are gauge invariant.  In writing \eref{eq:curlB}, we have dropped the mass term $(d\tW/dt)^2 {\bm A}_{\Acal}$ from the right-hand side.  It is numerically negligible since $d\tW/dt \sim H d\tW/d\ln x$ and the coherence length of the field $\lambda$ is much smaller than the Hubble scale $H^{-1}$.  
}
for the field $\Acal_{\mu}(x)$,  
\begin{align}\label{eq:curlB}
	{\bm \nabla} \times {\bm B}_{\Acal} = {\bm J}_{\Acal} + \dot{\bm E}_{\Acal} 
	\per
\end{align}
Combining these two formulas, we can show that 
\begin{align}\label{eq:EA}
	{\bm E}_{\Acal} = \frac{1}{\sigma_{\Acal}} {\bm \nabla} \times {\bm B}_{\Acal} - \frac{1}{\sigma_{\Acal}} {\bm J}_{\CME, \Acal} - {\bm v} \times {\bm B}_{\Acal}
\end{align}
where we have neglected the displacement current $\dot{\bm E}_{\Acal}$.  
This is justified in the magnetohydrodynamic (MHD) approximation \cite{Brandenburg:2004jv}, where $|\dot{\bm E}_{\Acal}|/|{\bm \nabla} \times {\bm B}_{\Acal}| \sim v/c \ll 1$. 
The term involving fluid velocity ${\bm v}$ does not contribute to the source term (\ref{eq:sources_3}) since ${\bm B}_{\Acal} \cdot {\bm v} \times {\bm B}_{\Acal}=0$. 

%==========
The chiral magnetic effect is the phenomenon whereby a magnetic field induces an electric current in a medium with a charge-weighted chiral asymmetry \cite{Vilenkin:1980fu}.  
By adapting the standard result for quantum electrodynamics \cite{Kharzeev:2013ffa} to our problem, the induced electric current can be written as 
\begin{align}\label{eq:JCMEA_def}
	{\bm J}_{\CME, \Acal} = \frac{g_\Acal^2}{2\pi^2} \mu_{5,\Acal} {\bm B}_{\Acal}
\end{align}
where $g_\Acal(t) \equiv g^\prime \cos \theta_W(t)$ is the effective gauge coupling for $\Acal_\mu$ and $\mu_{5,\Acal}$ is the charge-weighted chiral chemical potential.  
The corresponding charge-weighted chiral charge abundance is given by $\eta_{5,\Acal} = \mu_{5,\Acal}T^2/6s$.  
The chiral charge abundance is constructed from the abundances for the various SM particle species as 
\begin{align}\label{eq:eta5A_def}
	\eta_{5,\Acal} = \sum_{i} \Bigl[ - q_{u_L \Acal}^2 \eta_{u_L^i} - q_{d_L \Acal}^2 \eta_{d_L^i} - q_{\nu_L \Acal}^2 \eta_{\nu_L^i} - q_{e_L \Acal}^2 \eta_{e_L^i} + q_{u_R \Acal}^2 \eta_{u_R^i} + q_{d_R \Acal}^2 \eta_{d_R^i}+ q_{e_R \Acal}^2 \eta_{e_R^i} \Bigr]
\end{align}
where the sum runs over the three fermion families.  
The effective charges can be read off of the Lagrangian upon using the ansatz in \eref{eq:Ansatz}.  
These charges are found to be 
\begin{subequations}\label{eq:couplings}
\begin{align}
%%%
	q_{u_L \Acal}(t) & = y_{Q} + \frac{1}{2}\frac{ \tan \tW(t)}{\tan\tWz } 
	\\
%%%
	q_{d_L \Acal}(t) & = y_{Q} - \frac{1}{2}\frac{ \tan \tW(t)}{\tan\tWz } 
	\\
%%%
	q_{\nu_L \Acal}(t) & = y_{L} + \frac{1}{2}\frac{ \tan \tW(t)}{\tan\tWz } 
	\\
%%%
	q_{e_L \Acal}(t) & =y_{L} - \frac{1}{2}\frac{ \tan \tW(t)}{\tan\tWz } 
	\\
%%%
	q_{u_R \Acal}(t) & = y_{u_R} 
	\\
%%%
	q_{d_R \Acal}(t) & = y_{d_R} 
	\\
%%%
	q_{e_R \Acal}(t) & = y_{e_R} 
\end{align}
\end{subequations}
where $y$'s are the corresponding hypercharges.  

%==========
Finally, we put these pieces together.  
By combining \erefs{eq:EA}{eq:JCMEA_def}, we evaluate the electric field.  
This lets us express the source terms (\ref{eq:sources_3}) as 
\begin{subequations}\label{eq:sources_4}
\begin{align}
%%%
	\Scal_{\rm w}^{\rm bkg} & 
	= \Bigl[ - \frac{g^2}{2} \sin^2 \tW \Bigr] \Scal_{\rm BdB} 
	+ \Bigl[ \frac{g^2}{2} \frac{d\tW}{d\ln x} \sin 2 \tW \Bigr] \Scal_{\rm AB} 
	+ \Bigl[ \frac{g^2 g^{\prime 2}}{2} \sin^2 \tW \cos^2 \tW \Bigr] \gamma^{\CME} \eta_{5,\Acal} \\
%%%
	\Scal_{\rm y}^{\rm bkg} & 
	= \Bigl[ - g^{\prime 2} \cos^2 \tW \Bigr] \Scal_{\rm BdB} 
	+ \Bigl[ - g^{\prime 2}  \frac{d\tW}{d\ln x} \sin 2 \tW \Bigr] \Scal_{\rm AB} 
	+ \Bigl[ g^{\prime 4} \cos^4 \tW \Bigr] \gamma^{\CME} \eta_{5,\Acal} \\
%%%
	\Scal_{\rm yw}^{\rm bkg} & 
	= \Bigl[ - 2 g g^{\prime} \sin \tW \cos \tW \Bigr] \Scal_{\rm BdB} 
	+ \Bigl[ 2 g g^{\prime} \frac{d\tW}{d\ln x} \cos 2 \tW \Bigr] \Scal_{\rm AB} 
	+ \Bigl[ 2 g g^{\prime 3} \sin \tW \cos^3 \tW \Bigr] \gamma^{\CME} \eta_{5,\Acal} 
	\per
\end{align}
\end{subequations}
where we have used $d\tW/dt = H d\tW/d \ln x$ and defined 
\begin{subequations}\label{eq:source_defs}
\begin{align}
	\Scal_{\rm BdB}(t) & \equiv \frac{1/(4\pi)}{\pi \sigma_{\Acal} sT} \overline{ {\bm B}_{\Acal} \cdot {\bm \nabla} \times {\bm B}_{\Acal}} \\
	\Scal_{\rm AB}(t) & \equiv \frac{H}{8\pi^2 sT} \overline{ {\bm A}_{\Acal} \cdot {\bm B}_{\Acal} } \\
	\gamma^{\CME}(t) & \equiv \frac{12}{\pi^2} \frac{1}{(4\pi)^2} \frac{\overline{{\bm B}_{\Acal} \cdot {\bm B}_{\Acal}}}{\sigma_{\Acal} T^3}
	\per
\end{align}
\end{subequations}
Due to the volume averaging, the source terms are independent of the spatial coordinate.  
They depend upon the temporal coordinate through the entropy density $s$, the temperature $T$, the conductivity $\sigma_{\Acal}$, and the volume-averaged field products.  

%==========
Equation~(\ref{eq:sources_4}) is one of the main results of this paper.  
It should be compared with Eqs.~(2.53)~and~(2.60) of our earlier work \cite{Kamada:2016eeb}.  
To regain Eqs.~(2.53)~and~(2.60), we can take $\tW(t)$ to be a step function and set $d\tW/d\ln x= 0$.  
In the present calculation, we have generalized to an (as yet) arbitrary $\tW(t)$.  
As such, it is not necessary to treat the symmetric and broken phase cases separately, as we did in \rref{Kamada:2016eeb}.  
Rather, \eref{eq:sources_4} interpolates smoothly between the two solutions that we found previously.  
The term proportional to $d\tW/d\ln x$ was overlooked in previous studies, and we will see that it can provide an efficient source of baryon number.

%==================================
% ANALYTIC EQUILIBRIUM SOLUTION
%==================================
\section{Analytic equilibrium solution}\label{sec:Analytic}

%==========
Previous studies \cite{Giovannini:1997eg,Giovannini:1997gp,Anber:2015yca,Fujita:2016igl,Kamada:2016eeb} have shown that a helical hypermagnetic field in the symmetric phase of the EW plasma sources baryon number, which thereby competes against the washout of baryon number by EW sphalerons \cite{Kuzmin:1985mm}.  
Unlike the earlier work, in \sref{sec:Derivation}, we have taken a more careful treatment for the evolution of the magnetic field through the EW crossover, specifically allowing for a time-dependent weak mixing angle $\tW(t)$.  
By doing so, we have identified an additional source term in the kinetic equation for baryon number, namely the $(d\tW/d\ln x)\Scal_{\rm AB}$ term in \eref{eq:sources_4}.  
Here, we examine the evolution of the baryon asymmetry analytically with an emphasis on the effect of varying $\tW$.  

%==========
We derive the kinetic equation for baryon number by combining the the kinetic equations in \rref{Kamada:2016eeb} with the sources in \eref{eq:sources_4}.  
Denoting the baryon number-to-entropy ratio as $\eta_B$, its kinetic equation takes the form 
\begin{align}\label{eq:detaB_dx}
	\frac{d \eta_B}{dx} 
	& = \frac{3}{4} \bigl( g^2 + g^{\prime 2} \bigr) \Bigl[ \bigl( \cos 2 \tW - \cos 2 \tWz \bigr)\Scal_{\rm BdB}+ 2  \frac{d\tW}{d\ln x} \sin 2 \tW \Scal_{\rm AB} \Bigr]
	- (\text{washout terms})
\end{align}
where $\tW = \tW(t)$ is the time-dependent weak mixing angle.  
In the presence of a helical magnetic field, the terms containing $\Scal_{\rm BdB}$ and $\Scal_{\rm AB}$ (\ref{eq:source_defs}) become nonzero and source baryon number.  
In the symmetric phase, the weak mixing angle vanishes $\tW = 0$, and $\Scal_{\rm BdB}$ drives the growth of baryon number.  
During the EW crossover, $\tW$ begins to increase, and $\Scal_{\rm AB}$ contributes to the baryon-number growth.  
After the crossover, $\tW$ approaches its vacuum value, $\tan \tWz = g^{\prime} / g$, and both source terms become inactive; {\it i.e.}, their coefficients vanish.  
As we will see, the coefficient of the new source term $\Scal_{\rm AB}$ can vanish more slowly than the coefficient of $\Scal_{\rm BdB}$, and therefore the baryon asymmetry can be enhanced compared to previous calculations.  

%==========
The growth of baryon number is inhibited by several washout processes.  
These include the chiral magnetic effect, the EW sphaleron, and the electron spin-flip interaction, which comes into equilibrium below $T \simeq 80 \TeV$ and communicates baryon-number violation to the right-chiral electron \cite{Campbell:1992jd}.  
The equilibrium baryon asymmetry $\eta_{B,{\rm eq}}(t)$ is controlled by the slowest (least efficient) washout processes.  
For $T \gtrsim 145 \GeV$, the CME and spin-flip processes are slowest, and for $T \lesssim 145 \GeV$, the EW sphaleron is slowest.  
Thus, we calculate the equilibrium baryon number separately for these two periods below.  

%==========
At sufficiently high temperatures, $T \gtrsim 145 \GeV$, the EW sphaleron efficiently violates baryon number, and the equilibrium baryon asymmetry is controlled by a combination of the slower chiral magnetic effect and electron spin-flip interactions.  
The CME tends to deplete the charge-weighted chiral charge abundance $\eta_{5,\Acal}$ (\ref{eq:eta5A_def}), and the electron spin-flip interactions tend to equilibrate left- and right-chiral electron abundances.  
In this way, EW sphalerons violate baryon number among the left-chiral fermions, and the other washout processes communicate baryon-number violation to the right-chiral fermions.  
As in \rref{Kamada:2016eeb}, we calculate the equilibrium baryon asymmetry in the regime where all of the SM processes are in chemical equilibrium except for the CME and electron spin-flip interactions.\footnote{
This approach assumes that spin-flip interactions with the background Higgs condensate are in equilibrium.  At higher temperatures when the Higgs condensate has not yet developed, these interactions do not occur.  
In this regime, the baryon asymmetry can be calculated with Eqs.~(3.6) and (3.7) in \rref{Kamada:2016eeb}, but those formulas also agree with \eref{eq:etaBeq_1} below up to an $O(1)$ prefactor.  
It is known that this treatment during EW crossover gives $O(1-10\%)$ error in the estimate \cite{Khlebnikov:1996vj}, but here we neglect it. 
}  
We also require the four conserved charges to vanish; these are $(B/3-L_i)$ number and electromagnetic charge: $\eta_{B}/3 - \eta_{L_i} = \eta_{\rm em} = 0$.  
As discussed in \rref{Kamada:2016eeb}, the baryon asymmetry in equilibrium in this regime can be read off from the kinetic equation for the first-generation right-chiral electron abundance.  
Under these assumptions, it is reduced to
\begin{equation}
	\frac{d\eta_{e_R^1}}{dx}= g^{\prime 2} \left[ \cos^2 \tW \, \Scal_{\rm BdB}+ \frac{d\tW}{d\ln x} \sin 2 \tW \, \Scal_{\rm AB}-\frac{37}{11}g^{\prime 2} \cos^4 \tW \gamma^{\CME} \eta_B \right] - \frac{37}{11}\left( \frac{1}{2} \bigl( \gamma^{11}_{\Ehe}+ \gamma_{\Nhe}^{11} \bigr)+ \gamma_{\Ee}^{11}\right) \eta_B
	\per
\end{equation}
The transport coefficients $\gamma_{\Ehe}^{11}$, $\gamma_{\Nhe}^{11}$, and $\gamma_{\Ee}^{11}$ were defined in \rref{Kamada:2016eeb}.  
The equilibrium condition $d\eta_{e_R^1}/dx \approx 0$ gives the behavior of the baryon asymmetry in equilibrium, 
\begin{equation}\label{eq:etaBeq_1}
	\eta_B^{\rm eq} \approx \frac{11}{37}\frac{g^{\prime 2} \bigl( \cos^2 \tW \, \Scal_{\rm BdB} +\frac{d\tW}{d\ln x} \sin 2 \tW \, \Scal_{\rm AB} \bigr)}{\frac{1}{2} \bigl( \gamma_{\Ehe}^{11} + \gamma_{\Nhe}^{11} \bigr) + \gamma_{\Ee}^{11} + g^{\prime 4} \cos^4 \tW \gamma^{\CME} }
	\per
\end{equation} 
By taking $\tW = 0$ and $d\tW/d\ln x = 0$ we regain Eq.~(3.10) of \rref{Kamada:2016eeb}.
Notice how the equilibrium solution takes the form of $(\text{source})/(\text{washout})$, which expresses the balance between these two competing effects.  

%==========
At lower temperatures, $T \lesssim 145 \GeV$, the EW sphaleron rate becomes exponentially suppressed as the weak gauge boson masses grow, but nevertheless the sphaleron remains in equilibrium until $T \approx T_{\rm sph,fo} \simeq 130 \GeV$ \cite{DOnofrio:2014kta}.  
In this window, the EW sphaleron is the slowest washout process, and therefore it controls the equilibrium baryon asymmetry.  
Assuming that all of the SM processes are in equilibrium except for the EW sphaleron, the kinetic equation for baryon number (\ref{eq:detaB_dx}) reduces to 
\begin{align}\label{eq:detaB_dx_low}
	\frac{d \eta_B}{dx} = \frac{3}{4} \bigl( g^2 + g^{\prime 2} \bigr) \Bigl[  2 \frac{d\tW}{d\ln x} \sin 2 \tW \Scal_{\rm AB} \Bigr]
	-\frac{111}{34}\gamma_{\rm w,sph} \eta_B
\end{align} 
where $\gamma_{\rm w,sph}$ is the transport coefficient associated with the EW sphaleron process \cite{Kamada:2016eeb}.  
Here, we omit the term that includes $\Scal_{\rm BdB}$ since generally it is much smaller than the term with $\Scal_{\rm AB}$ at this period.  
The baryon asymmetry is well approximated by the equilibrium solution of \eref{eq:detaB_dx_low}.  
Solving $d\eta_B/dx \approx 0$ gives 
\begin{equation}\label{eq:etaBeq_2}
	\eta_B^{\rm eq} \approx \frac{17}{37} \frac{\bigl( g^2 + g^{\prime 2} \bigr) \frac{d\tW}{d\ln x} \sin 2 \tW \, \Scal_{\rm AB}}{\gamma_{\rm w,sph}}
	\per 
\end{equation}
This contribution to the baryon asymmetry is only present when $d\tW/d\ln x \neq 0$, and consequently it was overlooked in previous studies that did not treat the evolution of the magnetic field through the EW crossover so carefully.

%==========
Let us summarize the results of the preceding calculation.  
During the temperature window $80 \TeV \gtrsim T \gtrsim 130 \GeV$, all of the SM processes are in thermal equilibrium, including the electron spin-flip interaction and the EW sphaleron.  
In this regime, the baryon asymmetry is well approximated by 
\begin{align}\label{eq:etaBeq}
	\eta_{B}^{\rm eq} = \frac{11}{37} \frac{g^{\prime 2} \bigl( \cos^2 \tW \, \Scal_{\rm BdB} + \frac{d\tW}{d\ln x} \sin 2 \tW \, \Scal_{\rm AB} \bigr)}{\frac{1}{2} \bigl( \gamma_{\rm Ehe}^{11} + \gamma_{\rm \nu he}^{11} \bigr) + \gamma_{\rm Ee}^{11} + g^{\prime 4} \cos^4 \tW \gamma^{\CME} } + \frac{17}{37} \frac{\bigl( g^2 + g^{\prime 2} \bigr) \frac{d\tW}{d\ln x} \sin 2 \tW \, \Scal_{\rm AB}}{\gamma_{\rm w,sph}}
	\per  
\end{align}
At lower temperatures, $T \lesssim T_{\rm sph,fo} \simeq 130 \GeV$ the EW sphaleron has frozen out, and this calculation overestimates the baryon asymmetry.  
If the source terms are still active when $T < T_{\rm sph,fo}$, because the conversion from hypermagnetic field into electromagnetic field is very slow, then there can be a further enhancement of the baryon asymmetry.  
This is obtained by neglecting the washout term and directly integrating \eref{eq:detaB_dx} to find 
\begin{equation}\label{eq:etaBafo}
	\eta_B(x) \approx \eta_B(x_{\rm sph,fo})+\frac{3}{4}\bigl( g^2 + g^{\prime 2} \bigr)\int_{x_{\rm sph,fo}}^x dx'  \Bigl[ 2 \frac{d\tW}{d\ln x} \sin 2 \tW \Scal_{\rm AB} \Bigr], 
\end{equation}
where $x_{\rm sph,fo} = x(T_{\rm sph,fo})$ is the time of the EW sphaleron freeze-out. 
If the magnetic field conversion is sufficiently gradual, then $d\tW/d\ln x$ remains nonzero for a long time, and the baryon asymmetry can be enhanced by as much as $O(10-10^2)$, as we will see in the next section. 

%==================================
% RESULTANT BARYON ASYMMETRY EVOLUTION
%==================================
\section{Resultant baryon asymmetry evolution}\label{sec:Results}

%================
In this section, we present the quantitative results.  
We solve the kinetic equations now using the source terms that were derived in \sref{sec:Derivation}.  
However, we must first clarify a few additional assumptions.  

%================
Following \rref{Kamada:2016eeb}, we assume that the magnetic field is maximally helical and that its spectrum is peaked at the length scale $\lambda_B(t)$ where the field strength is $B_p(t)$.  
This allows us to estimate the volume-averaged magnetic field products, which appear in \eref{eq:source_defs}, as follows: 
\begin{subequations}\label{eq:B_products}
\begin{align}
	\overline{ {\bm B}_{\Acal} \cdot {\bm \nabla} \times {\bm B}_{\Acal}} & \approx \pm \frac{2\pi}{\lambda_B(t)} B_p(t)^2 \\
	\overline{ {\bm A}_{\Acal} \cdot {\bm B}_{\Acal} } & \approx \pm \frac{\lambda_B(t)}{2\pi} B_p(t)^2 \\
	\overline{ {\bm B}_{\Acal} \cdot {\bm B}_{\Acal} } & \approx B_p(t)^2
	\per 
\end{align}
\end{subequations}
The $\pm$ sign indicates the helicity of the magnetic field.  
Hereafter, we assume that the maximally helical magnetic field has a positive helicity [{\it i.e.}, the + signs in \eref{eq:B_products} are used].  
Flipping the sign of the helicity simply flips the sign of the resultant baryon asymmetry.  

%================
It is well known that a freely decaying, maximally helical magnetic field in a turbulent plasma experiences the inverse cascade evolution where power is transported from small scales to large ones \cite{FLM:373402, Pouquet:1976zz, Kahniashvili:2012uj}.  
As in \rref{Kamada:2016eeb}, we assume that the primordial magnetic field experiences the inverse cascade from a time well before the EW crossover until recombination, and afterward it evolves adiabatically (simply diluting with the cosmological expansion).  
Thus, we can relate the field strength and coherence length in the early Universe, $B_p$ and $\lambda_B$, to their values today, $B_0$ and $\lambda_0$, via the scaling laws 
\begin{align}\label{eq:inv_casc}
	B_p = \left( \frac{a}{a_0} \right)^{-2} \left( \frac{\tau}{\tau_{\rm rec}} \right)^{-1/3} B_0
	\qquad \text{and} \qquad
	\lambda_B = \left( \frac{a}{a_0} \right) \left( \frac{\tau}{\tau_{\rm rec}} \right)^{2/3} \lambda_0
\end{align}
where $a$ is the scale factor and $\tau$ is conformal time.  
These formulas apply when $\tau \leq \tau_{\rm rec}$ with $\tau_{\rm rec}$ the conformal time at recombination, and for later times, the factors of $(\tau/\tau_{\rm rec})$ must be removed to describe the adiabatic evolution of the magnetic field.  
Implicitly, the scaling law assumes that backreaction from the presence of particle/antiparticle asymmetries in the plasma is negligible, and we justify this assumption in \aref{app:back_reaction}.  
We also impose the constraint $\lambda_0/{\rm pc} = B_0/(10^{-14} \Gauss)$, which is expected to hold for causally generated magnetic fields that are processed on small scales by MHD turbulence \cite{Banerjee:2004df} (see also the discussion in \rref{Kamada:2016eeb}).  

\begin{figure}[t]
\begin{center}
\includegraphics[height=7cm]{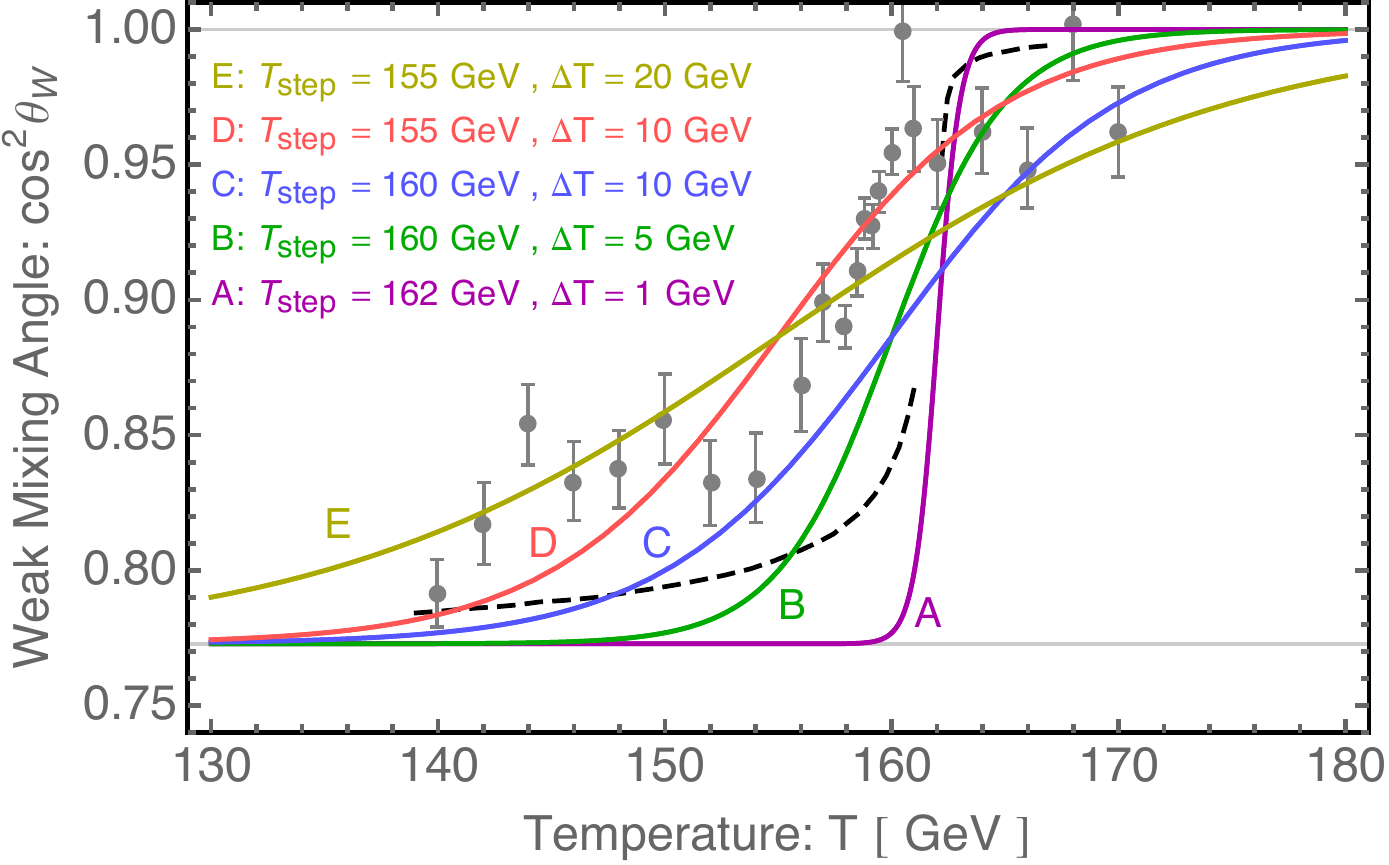}
\caption{
\label{fig:lattice_mixing}
The time-dependent weak mixing angle, expressed as $\cos^2 \tW(t)$.  
Results of numerical lattice simulations \cite{DOnofrio:2015mpa} appear as (gray) data points, and results of one-loop perturbative analytic calculations \cite{Kajantie:1996qd} appear as a (black) dashed line.  The other curves correspond to the ``smoothed step'' interpolating function from \eref{eq:smooth_step}, which we use for our analysis.  
}
\end{center}
\end{figure}

%================
The time-dependent weak mixing angle $\tW(t)$ has been calculated both analytically \cite{Kajantie:1996qd} and numerically \cite{DOnofrio:2015mpa}.  
We give these results in \fref{fig:lattice_mixing}.  
Evidently, the one-loop perturbative analytic calculation and the numerical lattice calculation agree only marginally.  
However, we can infer from both approaches that the weak mixing angle varies on a scale of $\Delta T \sim 10 \GeV$ during the EW crossover, which takes place at roughly $T \sim 160 \GeV$.  
Since the analytic calculation of \rref{Kajantie:1996qd} is only a one-loop result, the true behavior of $\tW(t)$ may differ when higher-order corrections are taken into account.  
Although the numerical lattice calculation is an all-orders calculation that includes nonperturbative effects, the error bars are still quite large.  
Since neither the analytic nor the numerical results for time dependence of the weak mixing angle appear more reliable, we will instead introduce a phenomenological parametrization for $\tW(t)$.  
Specifically, we write $\cos^2 \tW(t)$ as a smoothed step function, 
\begin{align}\label{eq:smooth_step}
	\cos^2 \tW(T) = \cos^2 \tWz + \frac{1-\cos^2 \tWz}{2} \left( 1 + \tanh \frac{T-T_{\rm step}}{\Delta T} \right)
	\com 
\end{align}
which interpolates between $\cos^2 \tWz = g^2 / (g^2 + g^{\prime 2}) \simeq 0.773$ at low temperature and $\cos^2 \tW = 1$ at high temperature.  
A few trial functions are also shown in \fref{fig:lattice_mixing}.  
It is straightforward to obtain $\tW$ in terms of the dimensionless temporal coordinate $x = M_0/T$. 

%============

%================
The conductivity of the SM plasma has been calculated in \rref{Arnold:2000dr}.  
In the symmetric phase at temperature $T \gg 100 \GeV$, they find the hypermagnetic conductivity to be $\sigma_{Y} \simeq 55 T$, and in the broken phase at temperature $T \sim 100 \GeV$, the electromagnetic conductivity is given by $\sigma_{\rm em} \sim 109 T$ (see also Ref.~\cite{Kamada:2016eeb}).  
The conductivity $\sigma_{\Acal}$ that appears in \eref{eq:JA} interpolates between these two limiting behaviors.  
However, for simplicity, we estimate the conductivity instead as $\sigma_{\Acal} = 100 T$ in both the symmetric and broken phases.  

\begin{figure}[t]
\begin{center}
\includegraphics[height=5.6cm]{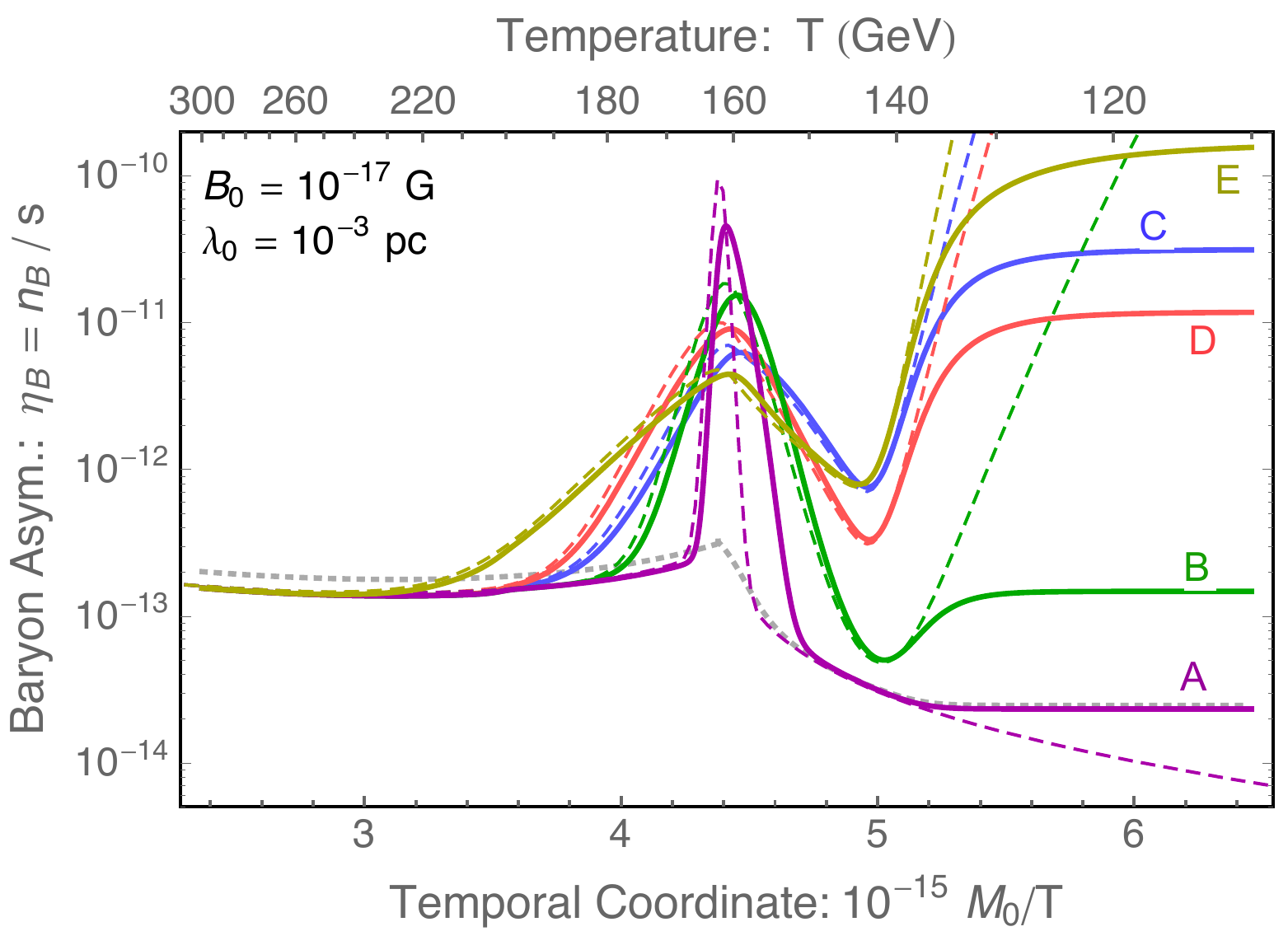} \hfill 
\includegraphics[height=5.6cm]{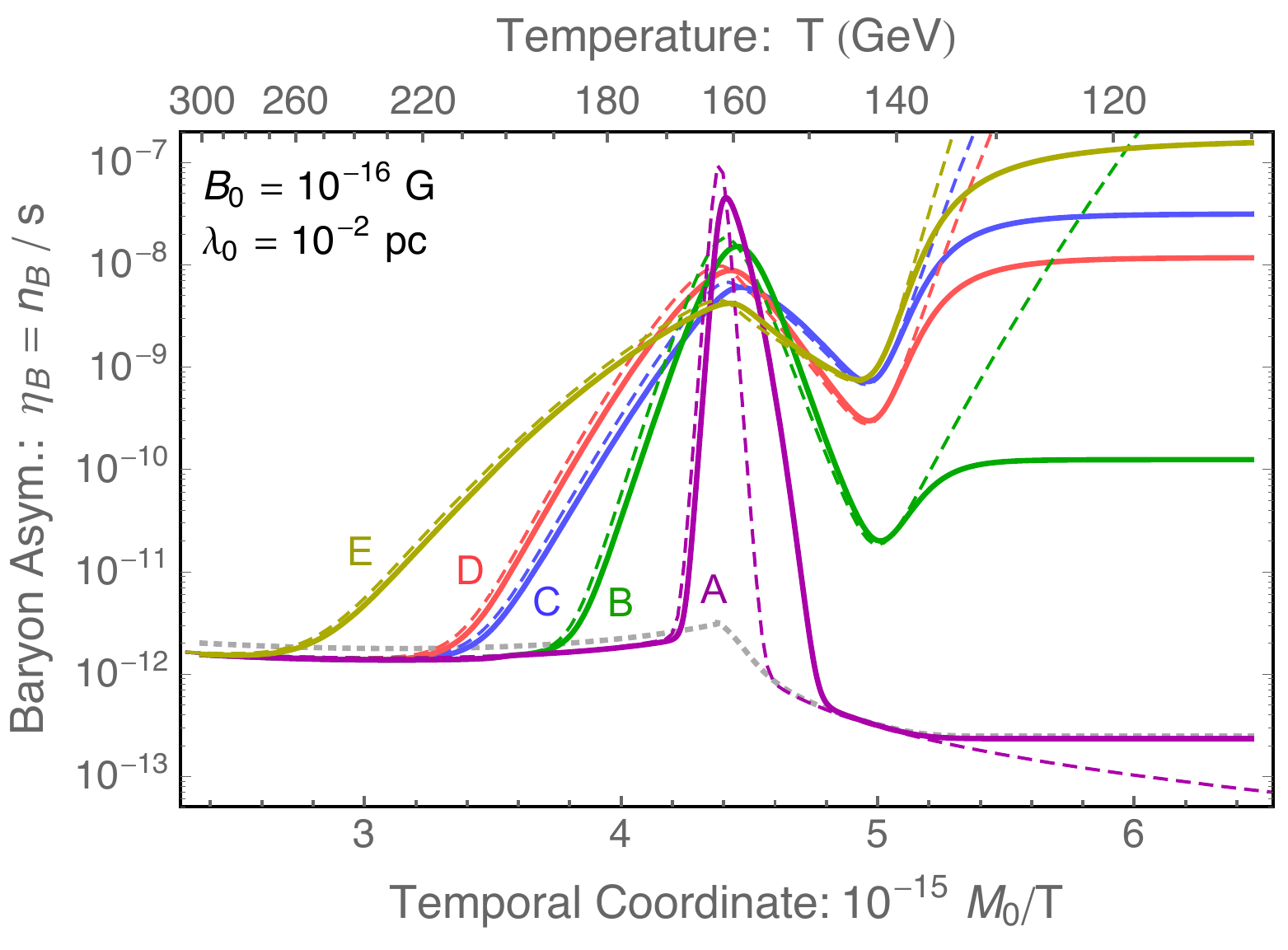} \\
\includegraphics[height=5.6cm]{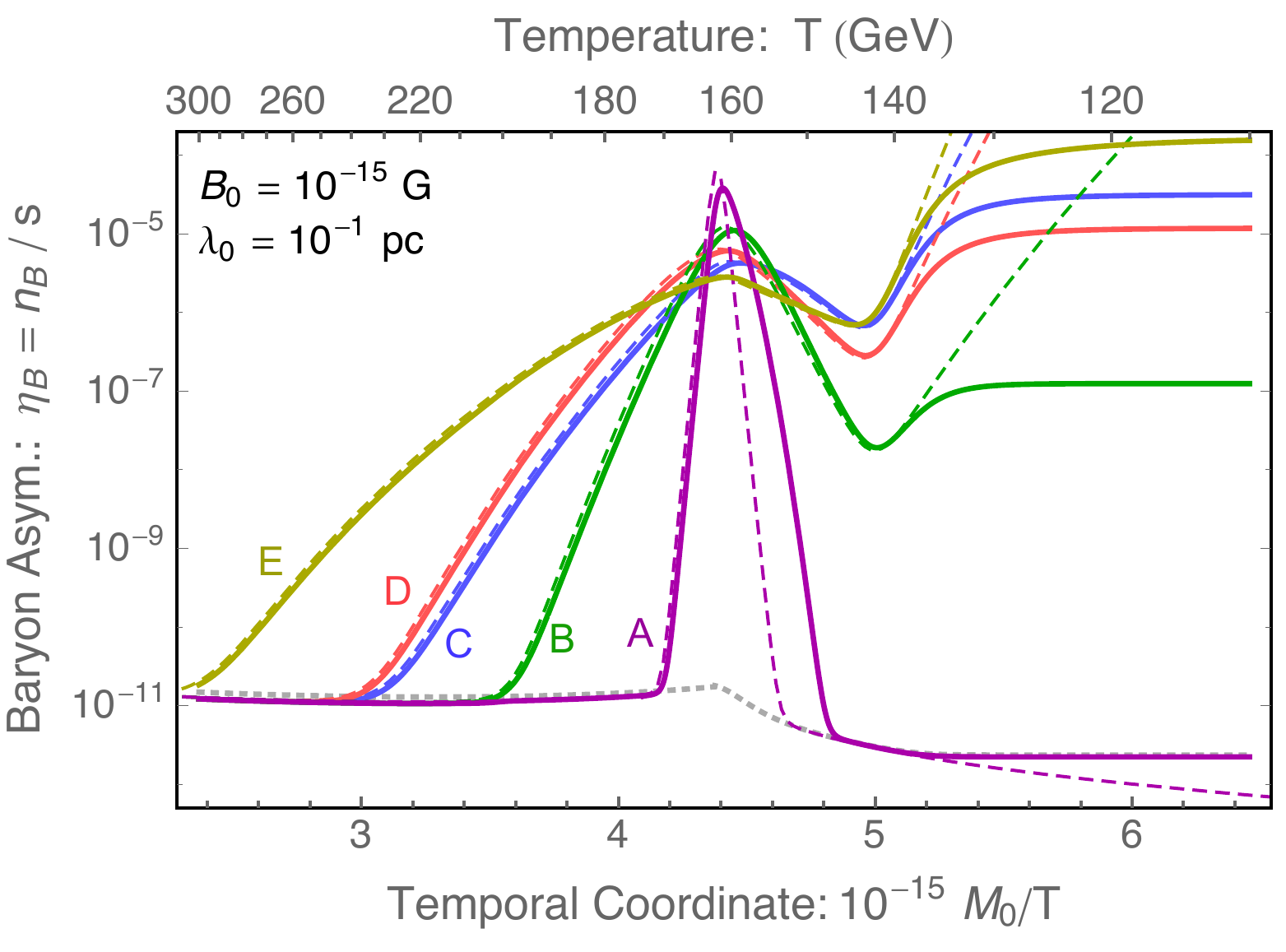} \hfill 
\includegraphics[height=5.6cm]{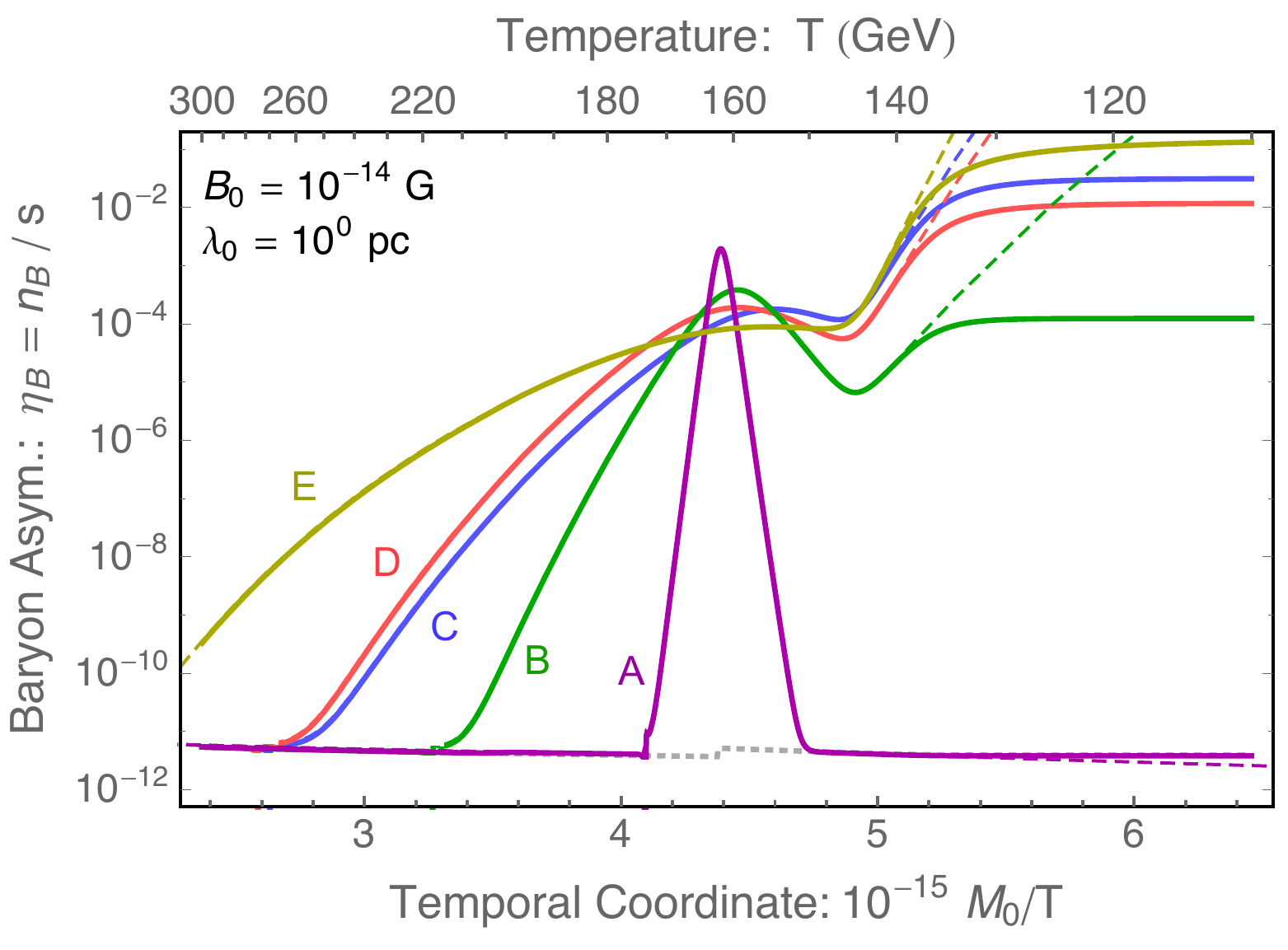} 
\caption{
\label{fig:etaB_versus_time}
Evolution of the baryon asymmetry $\eta_B$ during the EW crossover.  
The temporal coordinate is $x = T/H = M_0 / T$.  The four panels correspond to different values of the relic magnetic field strength $B_0$ and coherence length $\lambda_0$ today.  In each panel, the five pairs of colored curves correspond to the five parametrizations of $\tW(t)$ that appear in \fref{fig:lattice_mixing}.  The solid curves are the result of numerically solving the kinetic equations, and the dashed curves evaluate the formula in \eref{eq:etaBeq}.  The (gray) dotted curve corresponds to the calculation in \rref{Kamada:2016eeb}.  
}
\end{center}
\end{figure}

%================
Adopting \eref{eq:smooth_step} to model the time dependence of the weak mixing angle, we solve the kinetic equations \cite{Kamada:2016eeb} using the source terms in \eref{eq:sources_4}.  
The evolution of the baryon asymmetry during the EW crossover is shown in \fref{fig:etaB_versus_time}, where we compare the numerical solution with the analytic formula that appears in \eref{eq:etaBeq}. 
Evidently, the evolution of $\eta_B$ depends strongly on how the weak mixing angle evolves through the EW crossover; this behavior can be understood as follows.  

%================
Let us first consider the pair of (purple) curves which correspond to Parametrization A ($T_{\rm step}=162 \GeV, \Delta T=1 \GeV$) in \fref{fig:lattice_mixing}.  
In this case, the weak mixing angle quickly transitions between its asymptotic values at $T_{\rm step} = 162 \GeV$.  
The sudden change in $\tW$ implies an abrupt decrease in the helicity of the hypermagnetic field and a correspondingly large source of baryon number via the $\Scal_{\rm AB}$ term in \eref{eq:detaB_dx}.  
As predicted in \rref{Kamada:2016eeb}, the baryon number grows suddenly, but soon the hypermagnetic field is fully converted into an electromagnetic field, and the EW sphaleron, which remains in thermal equilibrium until $T \approx T_{\rm sph,fo} \simeq 130 \GeV$, is able to wash out the injection of baryon number.  
At temperatures $T \gtrsim 135 \GeV$, the analytic formula from \eref{eq:etaBeq} (dashed curve) matches the numerical result (solid curve) very well.  
After EW sphaleron freeze-out, $T \lesssim 130 \GeV$ the baryon number is fixed.  

%============

%================
The (gray) dotted curve in \fref{fig:etaB_versus_time} corresponds to the calculation of \rref{Kamada:2016eeb}, which assumed that the weak mixing angle changes abruptly and discontinuously at $T = 162 \GeV$ while $d\tW/d\ln x = 0$ at all times.  
The resultant relic baryon asymmetry agrees well with Model Parametrization A, which approximates the change in $\tW$ as a sudden but smooth step.  
The slight discrepancy between them can be traced to the factor of $\cos \tWz$ that arose in the calculation of \rref{Kamada:2016eeb} where ${\bm B}_{A}({\bm x},t_{\rm co}+\epsilon) = \cos \tWz \, {\bm B}_{Y}({\bm x},t_{\rm co}-\epsilon)$ was used to artificially match the hypermagnetic field into the electromagnetic field at the EW crossover.  

%================
For the models with a more gradual change in $\tW$, we see four distinct stages of evolution.  
First, $\eta_B$ begins to grow because $\tW$ (and hence $d\tW/d\ln x$) start to deviate from zero.  
This growth occurs earlier for the models of $\tW$ that have a broader step (larger $\Delta T$).  
The increase of $\eta_B$ continues until $T_{\rm step} \sim 160 \GeV$ where $d\tW/d\ln x$ peaks. 
The baryon asymmetry then decreases until $T \simeq 145 \GeV$ since the decrease of the source term with $d\tW/d\ln x$ is faster than that of the washout rate by the chiral magnetic effect and the electron spin-flip interaction.
At $T \simeq 145 \GeV$,  the EW sphaleron becomes the least efficient washout process.  
Afterward, $\eta_B$ grows as the EW sphaleron becomes less efficient at washout [$\gamma_{\rm w,sph}$ term in \eref{eq:etaBeq_2} decreases exponentially, much faster than the decay of the source term with $d\tW/d\ln x$]. 
This growth continues until $T_{\rm sph,fo} \simeq 130 \GeV$ where the EW sphaleron freezes out.  
The evolution of $\eta_B$ down to $T \simeq 135 \GeV$ is well described by the analytic solution in \eref{eq:etaBeq}, which appears as the dashed lines in \fref{fig:etaB_versus_time}.  
If the hypermagnetic field is not fully converted into an electromagnetic field by the time
 the EW sphaleron freezes out, there can be a continued growth of $\eta_B$, which is described by \eref{eq:etaBafo}.  
Eventually, the hypermagnetic field is fully converted into an electromagnetic field, and the relic baryon asymmetry is fixed.  
Practically, it is almost saturated\footnote{Note that the kinetic equations solved here neglect the effect of masses of the Higgs boson, weak bosons and top quarks and hence are not so reliable at low temperatures.  
However, since they do not contribute to the source term of the baryon number or the washout effects, we expect that there will not be a significant change of the baryon asymmetry and the numerical result at $T \sim 100 \GeV$ gives an appropriate estimate for the relic baryon asymmetry. } at $T\sim 100 \GeV.$  

\begin{figure}[t]
\begin{center}
\includegraphics[height=8cm]{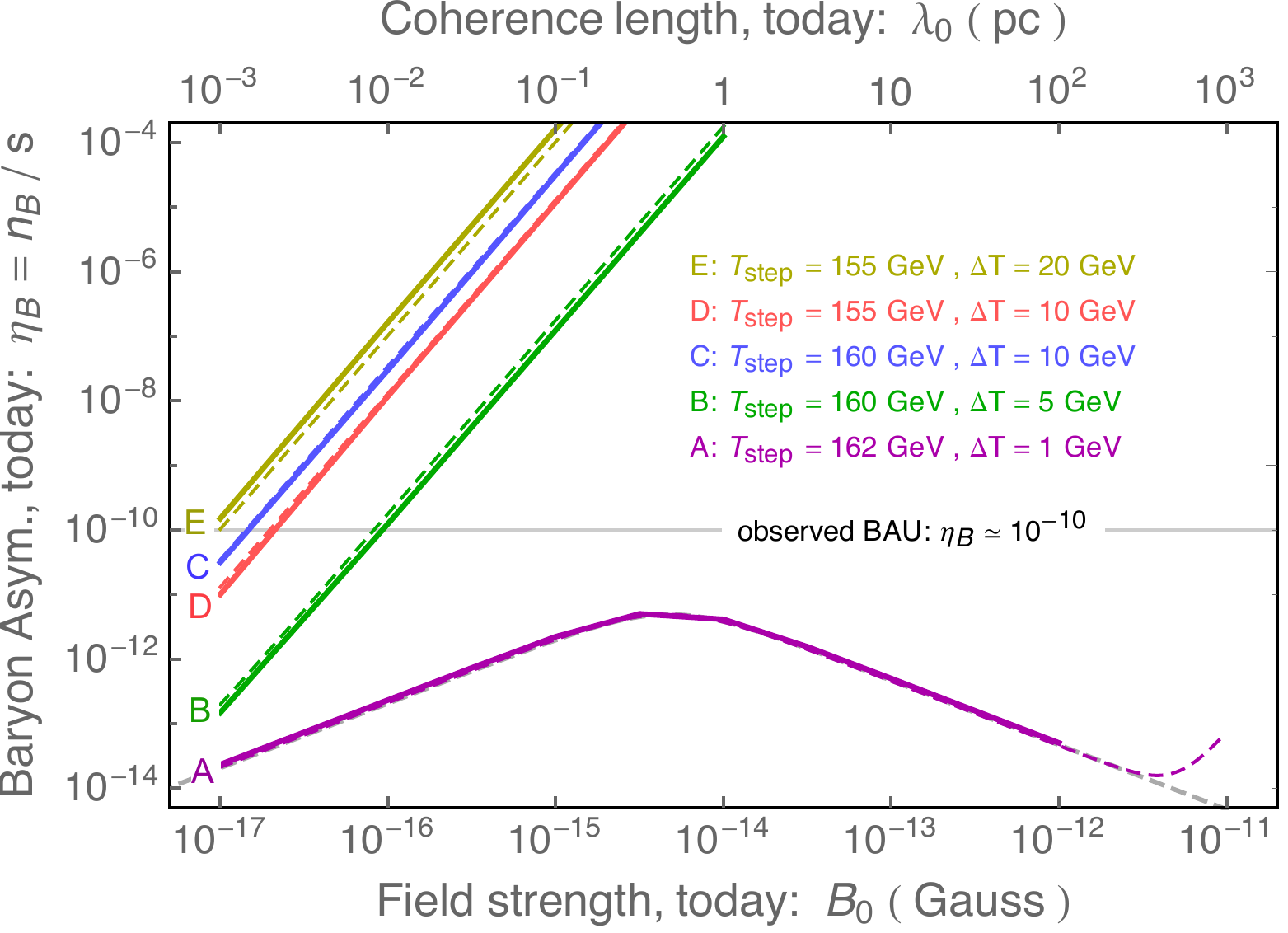}
\caption{
\label{fig:etaB_versus_B0}
The relic baryon asymmetry as a function of the relic magnetic field strength and coherence length today.  The five pairs of colored lines correspond to the different parametrizations of $\tW(t)$ in \fref{fig:lattice_mixing}: the solid lines show the result of numerical integration, $\eta_B(T=100\GeV)$, and the dashed lines show the analytic approximation (\ref{eq:etaBeq}) evaluated at $T = 135 \GeV$.  The (gray) dotted curve corresponds to the calculation in \rref{Kamada:2016eeb}.  
}
\end{center}
\end{figure}

%================
The relic baryon asymmetry [analytic formula \eref{eq:etaBeq} and numerical results] is shown in \fref{fig:etaB_versus_B0} as a function of the relic magnetic field strength today.  
It depends sensitively the evolution of the weak mixing angle $\tW(t)$.  
In Parametrization A where $\tW(t)$ rapidly interpolates between its asymptotic values, the relic baryon asymmetry always falls below the observed baryon asymmetry of the Universe $\eta_{B,{\rm obs}} \sim 10^{-10}$. 
In the other cases, we allow for a more gradual variation in $\tW(t)$, and the relic baryon asymmetry is much larger.  
The observed BAU is obtained for $B_0 \sim 10^{-16\sim 17} \Gauss$ and $\lambda_0 \sim 10^{-2 \sim 3} \pc$, depending on the evolution of $\tW$.  
For a weaker magnetic field, the baryon asymmetry is underpredicted, and an additional baryogenesis mechanism is required to explain cosmological observations.  
For a stronger magnetic field, the baryon asymmetry is over-predicted, and the model comes into tension with the observed baryon asymmetry.  
The relic BAU is particularly sensitive to the value of $\Delta T$, and by changing $\Delta T$ from just $5$ to $20 \GeV$, the relic BAU varies by up to 3 order of magnitude.  
Therefore, the accurate determination of $\tW(t)$ is necessary to reliably calculate the relic baryon asymmetry.  
Nevertheless, the qualitative behavior will be unchanged, and the problem of baryon overproduction will persist for large field strengths.  

%============

%================
Before we close this section, let us draw attention to the regime $\eta_B \gg 10^{-10}$.  
If the predicted baryon asymmetry is too large, then our calculation is unreliable.  
Specifically, in deriving the kinetic equations \cite{Kamada:2016eeb} we have assumed that $\mu_i/T \ll 1$ for the chemical potentials $\mu_i$ associated with each of the SM particle species.  
The corresponding abundance is calculated as $\eta = \mu T^2/(6s) \simeq (4 \times 10^{-3}) (\mu/T)$ with $s = (2\pi^2/45) g_{\ast S} T^3$ the entropy density and $g_{\ast S} \simeq 106.75$.  
Then, the condition $\mu_i/T \ll 1$ implies $\eta_i \ll 10^{-3}$.  
Consequently, the formula in \eref{eq:etaBeq} for the equilibrium baryon asymmetry cannot be trusted\footnote{One might wonder whether the conclusion of baryon-number overproduction can be avoided in the strong field regime where a more sophisticated calculation is required to accurately infer the late-time behavior of $\eta_B$.  While we cannot exclude this possibility outright, we cannot envisage any mechanism that would suppress $\eta_B$ back down to order $10^{-10}$.  } if $\eta_B \gg 10^{-3}$, but the calculation is certainly reliable for $\eta_B$ as large as $10^{-10}$.  
We discuss further in \aref{app:back_reaction} the reliability of our calculation in the large $\eta_B$ regime. 

%==================================
% AVOID OVER PRODUCTION
%==================================
\section{Avoiding baryon-number overproduction}\label{sec:Avoid}

%============
As we discussed in the Introduction, various blazar observations provide evidence for the existence of an intergalactic magnetic field with strength $B_0 \gtrsim 10^{-14\sim16} \Gauss$ and coherence length $\lambda_0 \gtrsim 1 \pc$.  
However, our calculations of the relic baryon asymmetry, which are summarized in \fref{fig:etaB_versus_B0}, imply that for such a strong PMF the BAU may be dramatically overproduced, $\eta_B \gg \eta_{B,{\rm obs}}$.  
Therefore, if there exists an intergalactic magnetic field at the level suggested by the blazar observations, and if it is a relic of the early Universe that became maximally helical before the EW crossover, then some amount of baryon-number generation is unavoidable due to Standard Model anomalies, and having calculated $\eta_B$ here, we identify a conflict between the inferred IGMF and the known baryon asymmetry of the Universe.  
In drawing this conclusion, we echo the earlier concerns of Fujita and Kamada~\cite{Fujita:2016igl}.  
In this section, we discuss a few ways to avoid this tension.

%================
It is possible to avoid the overproduction of baryon number by relaxing one (or more) of the assumptions that went into our analysis.  
These assumptions and possible ways out are cataloged below:
\begin{enumerate}
%%%
	\item[1. \, ]  We have assumed that the primordial hypermagnetic field is present in the symmetric phase of the EW plasma.  In the broken phase, the electromagnetic field, which has vectorlike interactions, does not contribute to the $(B+L)$ anomaly.  If the primordial magnetic field arises after the EW crossover has occurred ($T \lesssim 100 \GeV$), then there is no baryogenesis.  
%%%
	\item[2. \, ]  We have assumed that the magnetic field is maximally helical.  In this case, either the left- or right-circular polarization mode amplitude is dominant, and we can estimate the magnetic field products as in \eref{eq:B_products}.  Instead, if the magnetic field is nonhelical, then it does not source a global baryon number [$\Scal_{\rm BdB} = \Scal_{\rm AB} = 0$ in \eref{eq:detaB_dx}], and there is no baryogenesis.  

More generally, if the magnetic field is partially helical at the EW epoch, then the relic baryon asymmetry is suppressed by the helicity fraction.  
For this case, the calculation in \sref{sec:Results} must be modified as follows.  
For a nonhelical magnetic field, the inverse cascade scaling relations of \eref{eq:inv_casc} are replaced with the direct cascade scaling relations according to \rref{Kahniashvili:2012uj}, 
\begin{align}\label{eq:casc}
	B_p = \left( \frac{a}{a_0} \right)^{-2} \left( \frac{\tau}{\tau_{\rm rec}} \right)^{-1/2} B_0
	\qquad \text{and} \qquad
	\lambda_B = \left( \frac{a}{a_0} \right) \left( \frac{\tau}{\tau_{\rm rec}} \right)^{1/2} \lambda_0
\end{align}
for $\tau \leq \tau_{\rm rec}$. 
Let us denote the helicity fraction at conformal time $\tau$ by $\epsilon(\tau)$ and note that $0 \leq \epsilon \leq 1$.  
Then, the magnetic field products in \eref{eq:B_products} should be generalized to 
\begin{align}
	\overline{ {\bm B} \cdot {\bm \nabla} \times {\bm B}} \approx \pm \epsilon \frac{2\pi}{\lambda_B} B_p^2 
	\ , \qquad
	\overline{ {\bm A} \cdot {\bm B} } \approx \pm \epsilon \frac{\lambda_B}{2\pi} B_p^2 
	\ , \quad \text{and} \qquad
	\overline{ {\bm B} \cdot {\bm B} } \approx B_p^2
	\per 
\end{align}
Since comoving helicity is approximately conserved, $\Hcal = a(t)^3 \overline{{\bm A} \cdot {\bm B}}$, we see that the helicity fraction grows as $\epsilon(\tau) = (\tau/\tau_{\rm rec})^{1/2} \epsilon_0$ for $\tau \leq \tau_{\rm rec}$ where $\epsilon_0 \leq 1$ is the helicity fraction today.\footnote{We are considering the case where the helicity of the primordial (hyper)magnetic fields is extremely tiny, and hence it does not become maximally helical until today. }  
Since the relic baryon asymmetry is primarily controlled by $\Scal_{\rm AB} \propto \overline{ {\bm A} \cdot {\bm B} }$, we expect that the BAU is suppressed by a factor of
\begin{align}
	\frac{\Scal_{\rm AB}(\tau_{\EW})|_{\rm new}}{\Scal_{\rm AB}(\tau_{\EW})|_{\rm old}} 
	= \frac{\Bigl[ \left( \frac{\tau_{\EW}}{\tau_{\rm rec}} \right)^{1/2} \epsilon_0 \Bigr] \Bigl[ \left( \frac{\tau_{\EW}}{\tau_{\rm rec}} \right)^{1/2} \lambda_0 \Bigr] \Bigl[ \left( \frac{\tau_{\EW}}{\tau_{\rm rec}} \right)^{-1/2} B_0 \Bigr]^2}{\Bigl[ \left( \frac{\tau_{\EW}}{\tau_{\rm rec}} \right)^{2/3} \lambda_0 \Bigr] \Bigl[ \left( \frac{\tau_{\EW}}{\tau_{\rm rec}} \right)^{-1/3} B_0 \Bigr]^2}
	= \epsilon_0
\end{align}
compared to our previous estimates.  
Thus for a given $B_0$ and $\lambda_0$ that lead to baryon-number overproduction in the maximally helical case, it is possible to chose a $\epsilon_0 \ll 1$ such that the partially helical magnetic field reproduces the observed BAU.  
	\item[3. \, ]  We have assumed that the cosmological medium evolves adiabatically during the epoch between the EW crossover and today.  Consequently there is a conserved quantity $\eta_B = n_B / s$ with $n_B$ the number density of baryon number and $s$ the entropy density.  Instead, if there is an entropy injection after EW crossover, then $\eta_B$ will decrease, and baryon-number overproduction can be avoided with a sufficient amount of dilution.  However, the late-time entropy production also dilutes the magnetic field relative to the plasma energy density.  Hence, we expect that it is difficult to accommodate the observed BAU while simultaneously generating a strong enough IGMF to explain the blazar observations.  
\end{enumerate}
By relaxing any one of these assumptions, we can avoid the problem of baryon-number overproduction, but only Cases  2 and 3  are able to accommodate the observed BAU.  

%================
Finally, we have assumed that the coherence length is initially small enough that the magnetic field evolves subject to the turbulent motions of the cosmological plasma and the inverse cascade scaling behavior (\ref{eq:inv_casc}) is reached before the EW epoch.  
If instead the initial coherence length is much larger than the one determined by MHD turbulence, then the magnetic field evolves adiabatically at first and only enters the inverse cascade regime when the eddy scale catches up the coherence scale, which can be at a time after the EW crossover.  
It was shown in \rref{Fujita:2016igl} that the magnetic field strength is smaller for this initially adiabatic scenario than for the purely inverse cascade scenario.  
Therefore, one might expect that the resultant baryon asymmetry is suppressed, but as we see from the following estimates, this is not the case.  

%================
For the initially adiabatic scenario, the scaling relations (\ref{eq:inv_casc}) are replaced by 
\begin{align}\label{eq:late_inv_casc}
	B_p = \left( \frac{a}{a_0} \right)^{-2} \left( \frac{\tau_{\TS}}{\tau_{\rm rec}} \right)^{-1/3} B_0
	\qquad \text{and} \qquad
	\lambda_B = \left( \frac{a}{a_0} \right) \left( \frac{\tau_{\TS}}{\tau_{\rm rec}} \right)^{2/3} \lambda_0
\end{align}
where $\tau \leq \tau_{\TS}$ and $\tau_{\TS}$ denotes the conformal time when the inverse cascade scaling begins, which is assumed to be after the EW epoch, $\tau_{\EW} < \tau_{\TS}$. 
Consequently the source terms, $\Scal_{\rm BdB} \sim B_p^2 / \lambda_B$ and $\Scal_{\rm AB} \sim \lambda_B B_p^2$, are modified as 
\begin{subequations}
\begin{align}
	\frac{\Scal_{\rm BdB}(\tau_{\EW})|_{\rm new}}{\Scal_{\rm BdB}(\tau_{\EW})|_{\rm old}} 
	& = \frac{\Bigl[ \left( \frac{\tau_{\TS}}{\tau_{\rm rec}} \right)^{2/3} \lambda_0 \Bigr]^{-1} \Bigl[ \left( \frac{\tau_{\TS}}{\tau_{\rm rec}} \right)^{-1/3} B_0 \Bigr]^2}{\Bigl[ \left( \frac{\tau_{\EW}}{\tau_{\rm rec}} \right)^{2/3} \lambda_0 \Bigr]^{-1} \Bigl[ \left( \frac{\tau_{\EW}}{\tau_{\rm rec}} \right)^{-1/3} B_0 \Bigr]^2}
	= \left( \frac{\tau_{\EW}}{\tau_{\TS}} \right)^{4/3}
	\\
	\frac{\Scal_{\rm AB}(\tau_{\EW})|_{\rm new}}{\Scal_{\rm AB}(\tau_{\EW})|_{\rm old}} 
	& = \frac{\Bigl[ \left( \frac{\tau_{\TS}}{\tau_{\rm rec}} \right)^{2/3} \lambda_0 \Bigr] \Bigl[ \left( \frac{\tau_{\TS}}{\tau_{\rm rec}} \right)^{-1/3} B_0 \Bigr]^2}{\Bigl[ \left( \frac{\tau_{\EW}}{\tau_{\rm rec}} \right)^{2/3} \lambda_0 \Bigr] \Bigl[ \left( \frac{\tau_{\EW}}{\tau_{\rm rec}} \right)^{-1/3} B_0 \Bigr]^2}
	= 1
\end{align}
\end{subequations}
where we have used the scaling relations in \erefs{eq:inv_casc}{eq:late_inv_casc}
Indeed, $\Scal_{\rm BdB}$ is suppressed at the EW crossover by $(\tau_{\EW}/\tau_{\TS})^{4/3} < 1$, which is the origin of the suppression of the BAU in \rref{Fujita:2016igl}.  
On the contrary, $\Scal_{\rm AB}$ is unchanged for the same $B_0$ and $\lambda_0$. 
Since the main source of baryon overproduction at the EW crossover is $\Scal_{\rm AB}$, the problem cannot be avoided even in the initially adiabatic scenario. 
This also suggests that baryon overproduction is hardly avoided for the maximally helical magnetic fields with large correlation length $\lambda_0/{\rm pc} > B_0/(10^{-14}\Gauss)$, 
which are generated acausally and evolve fully adiabatically until today. 
We have seen that $\Scal_{\rm AB}$ is independent of the evolution of magnetic fields 
but only depends on $\lambda_0$ and $B_0$.  
Since $\Scal_{\rm AB}$ is proportional to $\lambda_0 B_0^2$, for larger correlation length,
larger $\Scal_{\rm AB}$ is obtained, which predicts baryon overproduction even in the case of 
larger correlation length with fully adiabatic evolution.

%==================================
% CONCLUSION
%==================================
\section{Conclusion}\label{sec:Conclusion}

%================
In this work we have studied the evolution of the baryon asymmetry through the EW crossover in the presence of a helical magnetic field.  
Building on earlier work, we have now taken into account the gradual conversion of the hypermagnetic field into an electromagnetic field during the crossover.  
This effect is described by the time-dependent weak mixing angle $\tW(t)$.  
Since a robust and accurate calculation of $\tW(t)$ is not available in the literature, we have studied a few phenomenological parametrizations, which appear in \fref{fig:lattice_mixing}.  
For each of these parametrizations, we solve a system of kinetic equations to determine the evolution of the baryon asymmetry during the EW crossover.  

%================
The main result, which appears in \fref{fig:etaB_versus_time}, reveals that a large injection of baryon number occurs when the hypermagnetic field is converted into an electromagnetic field.  
This is because the $(B+L)$ number is sourced by changes in hypermagnetic helicity via the Standard Model anomalies (\ref{eq:Bdot}), and the hypermagnetic helicity decreases significantly when the hypermagnetic field is converted into an electromagnetic field.  
If $\tW(t)$ is sufficiently slowly varying, as we expect from lattice simulations (\fref{fig:lattice_mixing}), then this baryon asymmetry is not fully washed out by EW sphalerons, and the relic baryon asymmetry can be greatly enhanced compared to previous calculations, which can be seen in \fref{fig:etaB_versus_B0}.  

%================
In this way, the observed baryon asymmetry of the Universe is obtained for a maximally helical magnetic field with positive helicity and present-day field strength and coherence length of $B_0 \sim 10^{-17\sim16} \Gauss$ and $\lambda_0 \sim 10^{-3\sim2} \pc$.  
A maximally helical primordial magnetic field is generated naturally in axion models of inflationary magnetogenesis (the predictions for its present strength are still under discussion, though; see recent works in Refs.~\cite{Fujita:2015iga, Adshead:2016iae}). 

%================
Various measurements of TeV blazars have begun to uncover evidence for the existence of an intergalactic magnetic field with strength $B_0 \gtrsim 10^{-14\sim16} \Gauss$.  
For such a strong magnetic field, our calculation implies that the baryon asymmetry can be overproduced by many orders of magnitude.  
Anticipating that future observations will provide firm evidence for the existence of a strong IGMF, we have assessed in \sref{sec:Avoid} various ways of avoiding baryon-number overproduction.  
For instance, the relic primordial magnetic field may be a subdominant component of the present intergalactic magnetic field.

%================
In closing, let us remark upon how the calculation could be extended and improved.  
As we have seen, the resultant baryon asymmetry is strongly dependent on how we parametrize the time dependence of the weak mixing angle $\tW(t)$ during the EW crossover.  
We have been forced to employ oversimplified parametrizations for $\tW(t)$, see \fref{fig:lattice_mixing}, which are motivated by the one-loop analytic calculation and the most recent numerical lattice simulations.  
In order to more accurately determine $\tW(t)$, we would encourage that the analytic calculations be extended beyond the one-loop order, and the precision of the numerical lattice calculations is improved.  
Of particular importance is the behavior of $\tW(t)$ at temperatures $T \lesssim 140 \GeV$, because at these temperatures the EW sphaleron goes out of equilibrium, and the baryon asymmetry is able to grow without washout.

%----------------------------------------------------------------
% Acknowledgements
%----------------------------------------------------------------
\quad \\
\noindent
{\bf Acknowledgments:} 
K.~K. acknowledges support from the DOE for this work under Grant No. DE-SC0013605.
A.~J.~L. is supported at the University of Chicago by the Kavli Institute for Cosmological Physics through Grant No. NSF PHY-1125897 and an endowment from the Kavli Foundation and its founder Fred Kavli.  
We are very grateful to Mikhail Shaposhnikov for emphasizing the time dependence of the weak mixing angle, which was a primary motivation for this work.  
We are also thankful to Yoshiyuki Inoue for pointing out new blazar constraints. 

%----------------------------------------------------------------
%----------------------------------------------------------------
%----------------------------------------------------------------
\appendix

%==================================
% Assess Back Reaction on Magnetic Field Evolution
%==================================
\section{Assess backreaction on magnetic field evolution}\label{app:back_reaction}

%================
In the present analysis (also Refs.~\cite{Fujita:2016igl,Kamada:2016eeb}), we have assumed that the background magnetic field evolves according to the inverse cascade scaling relation (\ref{eq:inv_casc}).  
The inverse cascade is observed in studies of freely decaying maximally helical magnetic fields subject to MHD turbulence.  
Such studies do not take into account the anomaly affects nor the presence of particle/antiparticle asymmetries in the plasma.  
In our calculation, these asymmetries can be large ($\eta \gg 10^{-10}$), and the reliability of the inverse cascade scaling relation becomes questionable.  
For instance, it is known that a large chiral asymmetry can lead to magnetic field growth or depletion through the chiral magnetic effect \cite{Boyarsky:2011uy}.  
In this Appendix, we assess the regime in which these effects can be neglected, which thereby justifies our use of the inverse cascade scaling law.  

%================
Let us begin with energetic considerations.  
The volume-averaged energy density of the magnetic field is given by 
\begin{align}
	\rho_B 
	& = \frac{1}{V} \int \! \ud^3 x \, \frac{1}{2} \Bigl( | {\bm E}_{\Acal}({\bm x},t) |^2 + | {\bm B}_{\Acal}({\bm x},t) |^2 \Bigr) 
	\approx \frac{1}{2} B_p(t)^2 \nn
	& \simeq \bigl( 20 \GeV^4 \bigr) \left( \frac{B_0}{10^{-14} \Gauss} \right)^2 \left( \frac{T}{100 \GeV} \right)^{14/3}
\end{align}
where we have used \eref{eq:inv_casc} to evaluate $B_p$ on the second line.  
The Helmholtz free energy density of the SM plasma at temperature $T$ is 
\begin{align}
	\mathcal{F} = - \frac{\pi^2}{90} g_{\ast} \,T^4 + \sum_{\rm species} O(\mu_i^2 T^2) + \cdots 
\end{align}
where $g_{\ast}(T)$ is the effective number of relativistic species.  
In the second term, we sum the chemical potentials $\mu_i$ of the various SM particle species.  
The dots indicate terms that are higher order in the small quantity $\mu_i/T$.  

%================
The anomaly allows us to increase $|\mu_i|$ at the expense of lowering $B_p$.  
When $\mu_i$ increases at the EW epoch due to the decaying hypermagnetic helicity, its growth is limited by energy conservation to satisfy $\Delta \Fcal < |\Delta \rho_B|$ if the system is in equilibrium.  
When expressed in terms of the corresponding charge abundance, $\eta = \mu T^2/(6s) \sim 10^{-3} \mu/T$, this condition becomes 
\begin{align}\label{eq:eta_bound}
	\eta \lesssim 10^{-6} \left( \frac{B_0}{10^{-14} \Gauss} \right) \left( \frac{T}{100 \GeV} \right)^{1/3}
	\per 
\end{align}
From these estimates, we conclude that the growth of the particle/antiparticle asymmetries at the EW epoch may have a negligible backreaction on the magnetic field evolution when \eref{eq:eta_bound} is satisfied.  
If \eref{eq:eta_bound} is violated, then energetic considerations suggest that it may not be justified to neglect the backreaction on the evolution of the magnetic field.  

%================
As a concrete source of the backreaction, we can consider the particle/antiparticle asymmetries, which affect the evolution of the magnetic field through the chiral magnetic effect.  
This can be seen as follows.  
Transcribing the relevant formulas from \sref{sec:Derivation}, the field equations are \begin{align}
	\frac{d}{dt} \bm B_{\Acal} = - {\bm \nabla} \times {\bm E}_{\Acal}
	\qquad \text{and} \qquad
	\frac{d}{dt} {\bm E}_{\Acal} = {\bm \nabla} \times {\bm B}_{\Acal} - {\bm J}_{\Acal} 
	\com
\end{align}
and the electric current ${\bm J}_{\Acal}$ is given by \eref{eq:JA}.  
Eliminating the electric field ${\bm E}_{\Acal}$ from these equations and using ${\bm \nabla} \cdot {\bm B}_{\Acal} = 0$, we obtain 
\begin{align}\label{eq:B_evolve}
	\frac{d}{dt} {\bm B}_{\Acal} = \Bigl[ 
	\frac{1}{\sigma_{\Acal}} \nabla^2 {\bm B}_{\Acal} 
	+ {\bm \nabla} \times \bigl( {\bm v} \times {\bm B}_{\Acal} \bigr) 
	\Bigr]_{\MHD} + \frac{g_{\Acal}^2}{2\pi^2} \frac{\mu_{5,\Acal}}{\sigma_{\Acal}} {\bm \nabla} \times {\bm B}_{\Acal} 
	\per
\end{align}
The terms in square brackets represent the standard MHD effects of magnetic diffusion and advection.  
Along with the Navier-Stokes equations, these terms lead the system to the inverse cascade scaling behavior.  
The remaining term corresponds to the chiral magnetic effect. 

%================
We move to Fourier space and decompose onto the circular polarization basis.  
Let $B_{\Acal}^{\pm}({\bm k},t)$ denote the amplitudes of the right- and left-circular polarization modes with wave vector ${\bm k}$ at time $t$.  
From \eref{eq:B_evolve} we see that the CME affects their evolution via 
\begin{align}\label{eq:CME_term}
	\frac{d}{dt} {B}_{\Acal}^{\pm}({\bm k},t) = \pm \frac{g_{\Acal}^2}{2\pi^2} \frac{\mu_{5,\Acal} k}{\sigma_{\Acal}} B_{\Acal}^{\pm}({\bm k},t) + \cdots 
\end{align}
where $k=|{\bm k}|$, and the dots indicate the MHD terms.  
If $\mu_{5,\Acal} > 0$ the right-circular polarization mode is amplified, while the left-circular polarization mode is suppressed.  
In this way, the growth of the charge-weighted chiral asymmetry $\mu_{5,\Acal}$ backreacts on the evolution of the magnetic field.  

%================
From \eref{eq:CME_term}, we can read off the time scale, $\tau = (2\pi^2 \sigma_{\Acal})/(g_{\Acal}^2 |\mu_{5,\Acal}| k)$.  
The effect of the CME on the magnetic field evolution can be neglected, while the age of the Universe $t_U \sim H^{-1}$ is much smaller than $\tau$.  
%The condition $t_U \ll \tau$ resolves to $|\mu_{5,\Acal}| \ll 2\pi^2 \sigma_{\Acal} H/g_{\Acal}^2k$.  
The spectrum of the magnetic field is peaked at the scale $k = 2\pi / \lambda_B(t)$.  
For these modes, the CME is negligible ($t_U \ll \tau$) as long as 
\begin{align}
	|\mu_{5,\Acal}| \ll \frac{\pi \sigma_{\Acal} H \lambda_B}{g_{\Acal}^2}
	\per
\end{align}
We estimate the right-hand side using \eref{eq:inv_casc} to calculate $\lambda_B$ at the EW epoch and using $g_{\Acal}^2 \approx g^{\prime 2} \simeq 0.1$.  
When expressed in terms of the charge abundance, $\eta_{5,\Acal} = \mu_{5,\Acal} T^2/(6s) \sim 10^{-3} \mu_{5,\Acal}/T$, the condition becomes
\begin{align}\label{eq:mu_bound}
	|\eta_{5,\Acal}| \ll 10^{-4} \left( \frac{\lambda_{0}}{{\rm pc}} \right) \left( \frac{T}{100 \GeV} \right)^{1/3} 
	\per
\end{align}
Typically, the chiral asymmetry is comparable in magnitude to the baryon asymmetry, $|\eta_{5,\Acal}| \sim |\eta_B|$, since they are both sourced by the helical magnetic field.  
From these estimates, we conclude that the growth of the particle/antiparticle asymmetries at the EW epoch has a negligible backreaction on the evolution of the magnetic field due to the chiral magnetic effect provided that $|\eta_{5,\Acal}| \ll 10^{-4} ( \lambda_{0} / {\rm pc} )$.  

%================
One might wonder whether the CME can become relevant after the crossover when $T$ is lower.  
For instance, at the time of recombination, $T \sim 0.1 \eV$, and \eref{eq:mu_bound} gives a stronger limit:  $|\mu_5|/T \lesssim 10^{-4} (\lambda_0/{\rm pc})$.  
However, this does not imply a corresponding limit on $|\mu_B|$.  
In the broken phase, baryon number is conserved, but chirality is largely washed out by spin-flip scatterings \cite{Pavlovic:2016mxq}.  
(A complete washout is avoided by the presence of the helical electromagnetic field.)  
Therefore, if the backreaction from CME is negligible at the EW crossover, it is also negligible afterward.  

%================
Let us close this section by comparing the bound in \eref{eq:eta_bound}, which is derived from the energetic argument, with \eref{eq:mu_bound}, which is derived from the CME argument.  
We make use of the relation $(B_0/10^{-14} \Gauss) = (\lambda_0 / {\rm pc})$, which is expected to be maintained [up to an $O(10)$ factor] for a causally generated PMF [see below \eref{eq:inv_casc}].  
Both bounds have the same scaling with temperature $T$.  
The bound derived from energetic considerations is stronger than the bound derived from the CME calculation by a factor of order $100$.  
This discrepancy is not necessarily inconsistent given the rough nature of our estimates.  
However, both arguments confirm that for $\eta_B \sim 10^{-10}$ we are justified in neglecting the backreaction on the magnetic field evolution.  

%----------------------------------------------------------------
% References
%----------------------------------------------------------------
\bibliographystyle{h-physrev5}
\bibliography{BAU_through_EW_Crossover}

\end{document}